\documentclass[prb,preprintnumbers,amsmath,amssymb]{revtex4}

\pdfoutput=1

\usepackage{graphicx}
\usepackage{bm}

\newcommand{\forget}[1]{}

\newcommand{\thalf}{\tfrac{1}{2}}
\newcommand{\half}{\frac{1}{2}}
\newcommand{\tquar}{\tfrac{1}{4}}

\newcommand{\sub}[1]{_{\text{#1}}}
\newcommand{\super}[1]{^{\text{#1}}}

\newcommand{\bigoh}[1]{O\left(#1\right)}

\newcommand{\real}{\operatorname{Re}}
\newcommand{\imag}{\operatorname{Im}}

\renewcommand{\arctan}{\operatorname{atan}}

\newcommand{\arctanh}{\operatorname{atanh}}

\newcommand{\sign}{\operatorname{sign}}

\newcommand{\Arg}{\operatorname{Arg}}

\newcommand{\modulo}[1]{\quad(\operatorname{mod} #1)}
\newcommand{\inlinemodulo}[1]{~(\operatorname{mod} #1)}

\renewcommand{\mod}{\operatorname{mod}}

\newcommand{\abs}[1]{\left\lvert #1 \right\rvert}

\newcommand{\set}[1]{{\{ {#1} \}}}

\newcommand{\goesto}{\rightarrow}

\newcommand{\bra}[1]{\left\langle #1 \right|\,}
\newcommand{\ket}[1]{\,\left| #1 \right\rangle}

\newcommand{\reals}{\mathbb{R}}
\newcommand{\complex}{\mathbb{C}}

\newcommand{\step}{\Theta}    
    
\newcommand{\kron}[2]{\delta_{#1, #2}}

\newcommand{\inv}[1]{#1^{-1}}

\newcommand{\conj}[1]{#1^*}

\newcommand{\definedby}{:=}
\newcommand{\defines}{=:}

\newcommand{\floor}[1]{\left\lfloor #1 \right\rfloor}

\newcommand{\perm}{\mathcal P}
\newcommand{\permq}{\mathcal Q}

\newcommand{\permsign}[1]{\lbrack #1 \rbrack}

\newcommand{\Spin}{S}

\newcommand{\Sx}{S\super{x}}
\newcommand{\Sy}{S\super{y}}
\newcommand{\Sz}{S\super{z}}

\newcommand{\ifwehave}[1]{\left[\operatorname{if}  #1 \right]}
\renewcommand{\therefore}{\Rightarrow}

\begin{document}

\title{Deformed strings in the Heisenberg model}

\author{Rob Hagemans}
\author{Jean-S\'ebastien Caux} 

\affiliation{Institute for Theoretical Physics, University of
  Amsterdam, 
  The Netherlands.}

\date{\today}

\begin{abstract}
We investigate solutions to the Bethe equations for the isotropic
$S = 1/2$ Heisenberg chain involving complex, string-like rapidity configurations
of arbitrary length.  Going beyond the traditional
string hypothesis of undeformed strings, we describe a general
procedure to construct eigenstates including strings with generic deformations, discuss
general features of these solutions, and
provide a number of explicit examples including complete solutions for all
wavefunctions of short chains. 
We finally investigate some singular cases and show from simple symmetry arguments that 
their contribution to zero-temperature correlation functions vanishes.
\end{abstract}

\maketitle

\tableofcontents




\section{Introduction}
The problem of diagonalizing the Hamiltonian of a generic quantum system is too difficult to be
carried out completely except in rather exceptional circumstances.  For noninteracting
models, this is easily done since multiparticle states are obtained from products of single
particle ones.  However, in the presence of interactions, finding the exact eigenstates
and energy eigenvalues becomes a problem of dimensionality equal to that of the Hilbert space.  A set of theories 
which stand out as an exception to this are so-called integrable models, the most
fundamental of which is the Heisenberg spin-$1/2$ chain \cite{Heisenberg1928},
\begin{align}
\label{eq:def_heisenberg_original}
	H = J \sum_{j=1}^N \left(S^x_j S^x_{j+1} + S^y_j S^y_{j+1} + (S^z_j S^z_{j+1} - 1/4)\right) - h \sum_{j=1}^N S^z_j
\end{align}
where $J$ is the magnetic exchange coupling constant and $h$ the external field.  
Throughout this paper, we will give results for the antiferromagnetic case $J = 1$
(the eigenstates are the same for $J < 0$, only their energies are reversed).
Since the $z$-component of the total spin commutes with $H$, it is
a good quantum number.  Starting from a reference state with all spins pointing up,
\begin{equation}
|0\rangle = \otimes_{j =1}^N |\uparrow \rangle_j,
\end{equation}
we can divide the Hilbert space into subsectors of fixed magnetization 
$\sigma = \frac{1}{N} \sum_j S^z_j$ spanned by the 
$\mbox{dim} (N,M) = \binom{N}{M}$
basis states with $M = N(1/2 - \sigma)$ overturned spins at
lattice positions $j_1, ..., j_M$,
\begin{equation}
|j_1, ..., j_M \rangle = S^-_{j_1} ... S^-_{j_M} |0 \rangle.
\end{equation}
The Schr\"odinger equation is solved by the Bethe Ansatz\cite{Bethe1931}, 
\begin{equation}
|\chi_M \rangle = \sum_{\{ j \}} \chi_M (j_1, ..., j_M) |j_1, ..., j_M \rangle, 
\end{equation}
\begin{equation}
\chi_M (j_1, ..., j_M) = \prod_{M \geq a > b \geq 1} sgn(j_a - j_b) 
\sum_{P} (-1)^{[P]} 
e^{i \sum_{a=1}^M k(\lambda_{P_a}) j_a + \frac{i}{2} \sum_{M \geq a > b \geq 1} sgn(j_a - j_b) \phi (\lambda_{P_a} - \lambda_{P_b})}
\label{eq:Bethe_Ansatz}
\end{equation}
with $\phi (\lambda) = 2 ~\mbox{atan} \lambda$ and $k(\lambda) = \pi - 2~\mbox{atan} 2\lambda$.
The energy of an eigenstate is $E = E_0 - h N \sigma$ with $E_0 = \sum_j \frac{-2J}{1 + 4\lambda_j^2}$.
The set of $M$ rapidities $\lambda$ are constrained by quantizing $H$ through the imposition of periodic boundary conditions,
yielding the Bethe equations
\begin{align}
\label{eq:bethe}
	\left[\frac{\lambda_\alpha - i/2}{\lambda_\alpha+i/2} \right]^N 
	= 
	\prod_\beta^{\beta\neq\alpha} \frac{\lambda_\alpha-\lambda_\beta-i}{\lambda_\alpha-\lambda_\beta+i},
\hspace{1cm} \alpha = 1,...,M.
\end{align}

The correspondence between the number of distinct solutions to (\ref{eq:bethe}) for given $M$ and the dimensionality
of the sub-Hilbert space is known as the completeness problem, and is a highly nontrivial
fact to verify.  To classify the eigenstates, the standard strategy is to consider the logarithm
of (\ref{eq:bethe}), 
\begin{align}
\label{eq:log_bethe_rap}
	2\arctan 2\lambda_\alpha = 2\pi \frac{J_\alpha}{N} + \frac{1}{N}\sum_\beta^{\beta\neq\alpha} 2\arctan (\lambda_\alpha-\lambda_\beta) 
        \hspace{1cm} \mbox{mod}~ {2\pi}
~,
\end{align}
introducing a set of quantum numbers ${J}$ (defined modulo $N$) which label the eigenstates
($J$ are half-odd integers for $N-M$ even, and integers for $N-M$ odd).  
Since the
Bethe wavefunctions formally vanish when two rapidities become equal, we could think that simply choosing
$M$ distinct quantum numbers among the set of $N$ allowed possibilities, which we can clearly do in 
$\mbox{dim} (N,M)$ ways, would allow us to reconstruct all the eigenstates in the subspace.  This, as
was known to Bethe himself, is too naive and simply fails 
\footnote{Interestingly, F. Bloch \cite{BlochZP61} studied the problem before Bethe but concluded that
the wavefunction form now known as the Bethe Ansatz yielded {\it too many} solutions to the eigenvalue
equation for $M = 2$.  This illustrates the crucial importance of counting states: had Bloch done it
correctly, we might not be speaking of Bethe's Ansatz.}.
The problem is that only some of the solutions to (\ref{eq:log_bethe_rap}) are in terms of real rapidities;
there also exist, as Bethe himself found, solutions involving groups of complex rapidities representing
bound states of magnons. 
This led Bethe to investigate this problem rather extensively in his original
paper, by attempting to explicitly construct all solutions.  He realized that complex rapidities typically
arrange themselves into regular patterns known as {\it strings}.  More importantly, he also realized that
there exist states in which these strings get deformed back into extra real solutions with coinciding
quantum numbers $J$ for which the wavefuntion is nonvanishing, meaning that the 'Pauli principle'
of allowing only single occupancy of the $J$ quantum numbers fails, and that counting states using these
is invalid \cite{Bethe1931,Essler1992}.  Bethe
proposed a scenario wherein all eigenstates could be obtained from real solutions complemented
by (deformed, possibly all the way back onto the real axis) string states, and showed that the counting of these solutions gives the correct
number of eigenstates.  

In the early 1970's, interest in general solutions to the Bethe equations was
revived by Takahashi's fundamental work on the thermodynamics of the isotropic Heisenberg chain
\cite{Takahashi1971}, making extensive use of the so-called string hypothesis in which only undeformed strings
are assumed to be present.  Gaudin \cite{GaudinPRL26}
further considered the gapped anisotropic chain, while Takahashi extended his study to the gapless 
anisotropic chain \cite{TakahashiPLA36}.  While these papers used the simplest form of the
string hypothesis, in which undeformed strings of arbitrary length were present, 
Takahashi's original treatment of the gapless case was shown by Johnson, McCoy and Lai
\cite{JohnsonPLA38} to yield an incorrect high-temperature expansion.  The reason for this 
discrepancy was that the states of a gapless anisotropic chain can only have strings of certain
allowed lengths, and upon restricting the original equations to this set, Takahashi and Suzuki 
obtained the correct thermodynamics \cite{TakahashiPLA41,TakahashiPTP48}.  As far as thermodynamics
are concerned, the string hypothesis as expounded in the above papers is understood to give
correct results as long as either the temperature or the magnetic field are not strictly vanishing
\cite{TsvelikWiegmann1983}.
Getting the thermodynamics right, however, in no way addresses the 'completeness' problem of the
Bethe Ansatz.  In fact, the successes of the Thermodynamic Bethe Ansatz using the string hypothesis
cast a long enduring shadow on the general awareness of Bethe's own understanding of the existence
and characteristics of deformed string solutions, which were 'rediscovered' only decades later.   
There now exists a large literature on this subject for the particular case of
the Heisenberg chains
\cite{Essler1992,FaddeevPLA85,FaddeevTakhtadzhyan1981,Woynarovich1982,Destri1982,Babelon1983,Vladimirov1984,KirillovJSM30,KirillovJSM36,KluemperZPB71,KluemperZPB75,IslerPLB319,JuettnerJPA26,JuettnerNPB430,TarasovIMRN1995,LanglandsLNP447,LanglandsCRM11,KirillovLiskova1997,Siddharthan1998,Ilakovac1999,Noh2000,FabriciusMcCoy2001,FabriciusMcCoy2001b,Fujita2003}.

The purpose of the present paper is to delve more deeply into the nature of Bethe eigenstates
which deviate significantly from the traditional string hypothesis.  As a starting point, we 
consider only the case of the isotropic antiferromagnet, but our main objective is to go
beyond the usual case of only 2 downturned spins typically considered in the existing literature.  
The motivation for such a study comes
in large part from recent work on dynamical correlation functions for large but finite chains
\cite{Caux2005,Caux2005b}, to which such states can in principle contribute, and for which the
string hypothesis is a good starting point but whose accuracy has to be confirmed state-by-state.  
More formally, it is our opinion that the completeness problem is not properly addressed by
arguments based exclusively on the string hypothesis:  we believe that
a true 'exact solution' of the Heisenberg chains requires providing an explicit scheme to
recover {\it all} the eigenstates from solutions to the Bethe equations, and that proofs of completeness have to go
hand in hand with a precise knowledge of what is being counted.  Our objective is to
make a step in this direction by proposing a scheme for states with deformed higher string-like complex rapidities.  

The paper is organized as follows.  In Section \ref{sec:Strings}, we set our notations and
discuss basic aspects of complex solutions to the Bethe equations.  In Section \ref{sec:deviations},
we derive the sets of equations for string deviations which are then used in Section \ref{sec:structure}
to obtain and discuss explicit solutions in various cases and limits.  Section \ref{sec:symmetric}
focuses on special degenerate cases and their form factors.  Finally, after the Conclusion, we give 
complete solutions to the Bethe equations for small chains in the Appendix.


\section{Strings in the Bethe equations}
\label{sec:Strings}

\subsection{Pairs}
\label{sec:pairs}
\label{sub:pairs}

It is easily seen that, if the set $\set\lambda$ is a solution to the Bethe equations, so is $\set{\conj\lambda}$. A stronger statement was proven by Vladimirov \cite{Vladimirov1986}, viz. that all solutions of the Bethe equations are self-conjugate, i.e. $\set\lambda = \set{\conj\lambda}$. 
As a consequence, complex roots of the Bethe equations always come in pairs of conjugate roots, $\set{\lambda_+,\lambda_-}$ where $\imag \lambda_+>0$ and $\lambda_- \definedby \conj\lambda_+$. 

Let us assume that the real parts of the various roots do not coincide, so $\real\lambda_j \neq \real\lambda_k$ if $j\neq k$, and that the pair is not centered on the origin. If the set $\set\lambda$ is symmetric, cases are possible where one or more pairs are in fact centered on the origin. Such cases are treated separately in section \ref{sec:symmetric}.

For the quantum numbers $\set{J_+,J_-}$ associated to the two conjugate roots, 
subtracting the Bethe equation associated to $\lambda_-$ from that of $\lambda_+$ gives
\begin{align}
	2\pi \frac{J_+-J_-}{N} &= 2\arctan 2\lambda - 2\arctan 2\conj\lambda 
	- \frac{2}{N} \left[ \arctan (\lambda-\conj\lambda) - \arctan (\conj\lambda-\lambda) \right] 
\notag\\
	&\quad
	- \frac{2}{N}\sum_k^{\lambda_k\neq\lambda_\pm}
	\left[ \arctan (\lambda-\lambda_k) - \arctan (\conj\lambda-\lambda_k) \right] \modulo{2\pi} 
	~.
\end{align}
We choose the branch cut of the logarithm such that $-\pi < \imag\ln z \leq \pi$. Then the branch cuts of the inverse tangent are such that
\begin{align}
	\arctan \conj z = 
	\left\{
	\begin{aligned}
		&\conj{\left(\arctan z\right)} + \pi &&\text{if $z \in ~]-i, -i\infty[$} \\
		&\conj{\left(\arctan z\right)} - \pi &&\text{if $z \in ~]i, i\infty[$} \\
		&\conj{\left(\arctan z\right)} &&\text{elsewhere.}
	\end{aligned}
	\right.
\end{align}
Taking the real part of the difference equation, taking into account the above branch cut and the fact that $\real(\arctan\lambda - \arctan\conj\lambda) = 0$ if $\real\lambda \neq 0$, and that we work modulo $\pi$, the real part of the equation becomes 
\begin{align}
\label{eq:wide_narrow}
	J_- - J_+ &= 
	\left\{
	\begin{aligned}
 	&1	&\text{if $\imag{\lambda_+} > \thalf$} 	\\
	&0 	&\text{if $\imag{\lambda_+} < \thalf$}
	\end{aligned}
	\right.
~.
\end{align}
In other words, there are two kinds of pairs: {\em narrow pairs} (also called {\em close pairs}), which are separated in the imaginary direction by less than $i$, and whose quantum numbers are equal; and {\em wide pairs}, which are separated by more than $i$ and whose quantum numbers differ by one, where the higher quantum number is associated to the root in the negative half-plane. This distinction is found in e.g. Destri and L\"owenstein \cite{Destri1982} and Babelon et al. \cite{Babelon1983}.
If $\imag{\lambda_+}=\half$, the Bethe equations become singular; we will study this important limit in more detail in the following sections.

The fact that solutions exist with repeated quantum numbers  makes the counting of allowed states very complicated. This problem is addressed by the {\em string hypothesis} which, among other things, introduces a new type of quantum number meant to be strictly non-repeating.


\subsection{The string hypothesis}

If a root $\lambda$ is complex with positive imaginary part and finite real part, the factor in the left-hand side of the Bethe equation
(\ref{eq:bethe}) has a norm less than unity. This implies that for large $N$, the left-hand side will vanish exponentially. Therefore, the right-hand side must vanish exponentially as well. Likewise, if the imaginary part is negative, both sides must diverge. 
The only way for a factor on the right-hand side to vanish for fixed $M$ as $N \rightarrow \infty$ 
is if there exists a root at a distance close to $i$ below the root under consideration, where the difference is close to $i$. Since this is true for every complex root we choose on the left-hand side, all non-real roots should be arranged in strings of various length, in which the roots that make up the string are spaced by distances close to $i$. This is a loose statement of 
what is known as the {\em string hypothesis}.

The $M$ roots of the Bethe equations are thus partitioned in a configuration of strings, 
where a $j$-string is a group of $j$ roots such that
\begin{align}
\label{eq:def_string}
	\lambda^j_{\alpha a} &= \lambda^j_\alpha + \tfrac{i}{2}\left(j +1 -2a\right) + d^j_{\alpha a} 
&
&\text{with $a \in \set{1, \ldots, j}$.}
\end{align}
Here $j$ is the string length, $\lambda^j_\alpha$ the string center and $d^j_{\alpha a} \equiv \epsilon^j_{\alpha a} + i \delta^j_{\alpha a}$ the  deviation, with $\epsilon^j_{\alpha a}, \delta^j_{\alpha a}\in\reals$. Furthermore, self-conjugacy dictates $d^j_{\alpha, a} = \conj{[ d^j_{\alpha, j+1-a}]}$.
Note that every self-conjugate configuration of roots can be given in terms of a set of strings as above albeit with arbitrarily 
large deviations. The string hypothesis assumes that all deviations vanish in the thermodynamic limit.

Using the parametrisation \eqref{eq:def_string}, let us now make our reasoning more precise. Consider the Bethe equations 
(\ref{eq:bethe}) for a root $\lambda^j_{\alpha a}$. Let $A^j_\alpha$ be the 
part of the product on the right side pertaining to roots on other strings. 
We assume that $A^j_\alpha$ is of order unity. 
Furthermore, the quotient on the left-hand side is denoted $z^j_{\alpha a}\in\complex$. We consider a root with positive imaginary part, so that $\abs{z^j_{\alpha a}}<1$.  The Bethe equation can be written
\begin{align}
\left[z^j_{\alpha a}\right]^N 
=
A^j_\alpha \prod_{b\neq a} 
	\frac
		{d^j_{\alpha a}-d^j_{\alpha b}+i(b-a-1)}
		{d^j_{\alpha a}-d^j_{\alpha b}+i(b-a+1)} 
	~.
\end{align}
Our parametrisation is such that $\imag\lambda^j_{\alpha a} > \imag\lambda^j_{\alpha a+1}$. Then, the positive-imaginary roots have $1 \leq a \leq \floor{j/2}$. 

Let us start at $a=1$. Since this is the top root, 
\begin{align}
\left[z^j_{\alpha a}\right]^N 
=
A^j_\alpha 
	\frac
		{d^j_{\alpha 1}-d^j_{\alpha 2}}
		{d^j_{\alpha 1}-d^j_{\alpha 2}+2i} 
	\frac
		{d^j_{\alpha 1}-d^j_{\alpha 3} + i}
		{d^j_{\alpha 1}-d^j_{\alpha 3} + 3i} 
	\cdots
~.
\end{align}
We now assume that the differences of $\delta$s are small enough for all but the first of the quotients on the right-hand side to be of order unity. Thus we have 
\begin{align}
\label{eq:toproot}
\abs{d^j_{\alpha 1}-d^j_{\alpha 2}} \propto 
\abs{z^j_{\alpha 1}}^N
~,
\end{align}
i.e., the difference must vanish exponentially with $N$. 

Now consider the next root, $a=2$. This time, 
\begin{align}
\left[z^j_{\alpha a}\right]^N 
=
A^j_\alpha 
	\frac
		{d^j_{\alpha 2}-d^j_{\alpha 1} - 2i}
		{d^j_{\alpha 2}-d^j_{\alpha 1} } 
	\frac
		{d^j_{\alpha 2}-d^j_{\alpha 3} }
		{d^j_{\alpha 2}-d^j_{\alpha 3} + 2i} 
	\frac
		{d^j_{\alpha 2}-d^j_{\alpha 4} + i}
		{d^j_{\alpha 2}-d^j_{\alpha 4} + 3i} 
	\cdots
~,
\end{align}
and therefore we conclude
\begin{align}
\abs{\frac{d^j_{\alpha 2}-d^j_{\alpha 3}} {d^j_{\alpha 2}-d^j_{\alpha 1}}} \propto 
\abs{z^j_{\alpha 2}}^N
~.
\end{align}
Multiplying by equation \eqref{eq:toproot} gives
\begin{align}
\abs{d^j_{\alpha 2}-d^j_{\alpha 3}} \propto 
\abs{z^j_{\alpha 2} z^j_{\alpha 1}}^N
~.
\end{align}
Continuing this reasoning until $a = \floor{j/2}$, we conclude that all differences of deviations vanish exponentially with $N$, while 
$\abs{d^j_{\alpha, a}-d^j_{\alpha, a+1}} \gg \abs{d^j_{\alpha, a+1}-d^j_{\alpha, a+2}}$
~.
Furthermore, we know that for $j$ even, $d^j_{\alpha, j/2} = -d^j_{\alpha, j/2+1}$ and $\real d^j_{\alpha, j/2}=0$; while for $j$ odd,  $d^j_{\alpha, \floor{j/2}} = 0$. Thus we conclude that all $d^j_{\alpha a}$ vanish exponentially with $N$, and their successive difference 
ratios as well, such that
$\abs{d^j_{\alpha, a}} \gg \abs{d^j_{\alpha, a+1}}$
~.
Note that during this derivation we have made a number of important assumptions, mainly that 
the factor $A^j_\alpha$ is of order unity and that the differences of consecutive $\delta$s are small. These assumptions do not necessarily hold at the same time: for instance, if many roots lie close to each other the derivation does not hold. In such cases, deviations may vanish more slowly. 
A more precise formulation of the string hypothesis is thus, that the deviations from string configurations as defined in \eqref{eq:def_string}, vanish exponentially with $N$ if $N$ is made large while all other parameters are kept constant.

Rewriting the Bethe equations under the assumption that the deviations vanish leads to the Bethe--Takahashi equations\cite{Takahashi1971}
\begin{equation}
\label{eq:log_bethe_takahashi}
2\arctan \frac{2\lambda^j_\alpha}{j}  = 2\pi \frac{I^j_\alpha}{N}  + \frac{1}{N}\sum_{k=1}^{N_s} \sum_{\beta=1}^{M_k} \Theta_{jk}(\lambda^j_\alpha - \lambda^k_\beta)\modulo{2\pi}
~,
\end{equation}
with 
\begin{equation}
\Theta_{jk} (\lambda) \definedby 2(1-\delta_{jk}) \arctan \frac{2\lambda}{\abs{j-k}} + 4\arctan\frac{2\lambda}{\abs{j-k}+2}
	+ \cdots + 4\arctan\frac{2\lambda}{j+k-2} + 2\arctan\frac{2\lambda}{j+k}
	~.
\end{equation}
where the $M$ roots are partitioned into $M_j$ strings of length $j$ such that $\sum_j j M_j = M$.  All positive
integer $j$ are in principle allowed.  For $N$ even, $I^j_{\alpha}$ are half-odd integers for $M_j$ even, and
integers for $M_j$ odd.  The string hypothesis assumes that these equations have $\mbox{dim} (N,M)$ distinct
solutions in terms of sets of real rapidities (allowing infinite ones).


\subsection{Completeness}

The Heisenberg Hamiltonian commutes with the total spin operator $\Spin^2\sub{tot} \definedby (\Sx\sub{tot})^2+ (\Sy\sub{tot})^2+ (\Sz\sub{tot})^2$ as well as with the total spin-z operator $\Sz\sub{tot} \definedby \sum_{j=1}^N \Sz_j$. States with $s=N/2-M$, where $s$ is the total-spin quantum number such that the eigenvalue of $\Spin^2\sub{tot}$ is $s(s+1)$, and $N/2-M$ is the $\Sz\sub{tot}$ eigenvalue, are heighest-weight states. From these states, we can generate lower-weight states by repeatedly applying the total-spin lowering operator $\Spin^-_0$. This corresponds to adding a particle with momentum $k=0$ and therefore rapidity $\lambda=\infty$. In particular, looking at a fixed-$M$ subspace with $M\leq N/2$, all solutions in the $(M-1)$-subspace are repeated with an extra infinite rapidity added. This means that in every $M$-subspace we only have to solve for the $\binom{N}{M} - \binom{N}{M-1}$ solutions of highest weight.

It has been proposed \cite{Bethe1931, Takahashi1971, Gaudin1972, FaddeevTakhtadzhyan1981} that in the Heisenberg model, 
the Bethe--Takahashi quantum numbers of highest-weight states should be non-coinciding and bounded by
\begin{align}
\label{eq:quantum_number_bound}
	\abs{I^j_{\alpha}} < I^j_\infty \definedby \half\Bigl[N+1-\sum_{k\geq 1} M_k(2\min(n_j, n_k)-\kron{j}{k}) \Bigr] 
	~.
\end{align}
One can prove \cite{Bethe1931,TakahashiBOOK} that the number of sets of quantum numbers that satisfy this constraint is 
$\binom{N}{M} - \binom{N}{M-1}$, exactly the number of highest-weight solutions with $M$ down spins. Therefore, provided that for each of 
 these quantum number choices a solution exists, is unique, and leads to an admissible solution of the Schr{\"o}dinger equation, 
 the Bethe Ansatz is complete.
 
In the analytically solvable case $M=2$, it is known that indeed, for each of these quantum numbers a unique, admissible solution exists. However, some of the quantum numbers associated to two-strings do not lead to complex-pair solutions, as narrow pairs get narrower and eventually merge and split back onto the horizontal axis.
Instead, in these cases, there are extra solutions with real roots. These extra real solutions can be connected one-to-one with the missing string solutions by way of the Bethe quantum numbers.




\section{String deviations}
\label{sec:deviations}

In this section we will derive equations for the exact string deviations. Determining these deviations is
therefore completely equivalent to determining the exact roots of the Bethe equations in the complex plane. 
We will first establish the method by considering two- and three-strings, and then generalise to arbitrary string lengths.


\subsection{Deviated two-strings}
The Bethe--Takahashi equations \eqref{eq:log_bethe_takahashi} are found from the sum of the logarithmic Bethe equations, absorbing all contributions from branch cut crossings in the Bethe--Takahashi quantum numbers. To find the relation of these 
with the Bethe quantum numbers, let us carefully redo the derivation. We concentrate on the isotropic chain. Similar but more complicated derivations can be given for the anisotropic chain.

We use the relation
\begin{align}
	\arctan (a+ib) + \arctan (a-ib) 
&= \xi(a, 1+b) + \xi(a, 1-b)
&\text{for $a,b \in\reals$,}
\end{align}
where we defined
\begin{align}
	\xi(\epsilon,\delta)\definedby
	\frac{1}{2i}\left[\ln (\delta+ i\epsilon) - \ln(\delta-i\epsilon)\right] 
	= \arctan \frac\epsilon\delta + \pi\, \step(-\delta) \sign \epsilon 
\end{align}	
We use $\sign 0 = 0$ and $\step$ is the Heaviside step function with $\step(0) = \thalf$. Note that $\xi(\epsilon, \delta) = \Arg (\delta+i\epsilon)$ for $\epsilon\neq0$, and $\xi(0, \delta)=0$.
A useful feature of $\xi(\epsilon,\delta)$ is that it is continuous on the line $\delta=0$. Therefore the value for zero deviations is well-defined, $\xi(\epsilon, 0) = \frac{\pi}{2}\sign\epsilon$ and we need not keep track of the sign of $\delta$ in the limit.

Deviated two-strings are parametrised as $\lambda^{(2)}_{\alpha\pm} \definedby \lambda^{(2)}_\alpha \pm i(1+2\delta^{(2)}_\alpha)/2$ where $\lambda^{(2)}_\alpha$ and $\delta^{(2)}_\alpha$ are real numbers. The sum of the (log-) Bethe equations for $\lambda^{(2)}_{\alpha\pm}$ equals, 
\begin{align}
\label{eq:two_string_with_deviance}
	&\xi({\lambda},{1+\delta}) + \xi ({\lambda}, {-\delta}) = 
	\pi\frac{J_+ + J_-}{N}
	+ \frac{1}{N}\sum^{k=1}_\beta \left[ \xi({\lambda-\lambda^k_\beta},{\delta+\tfrac32}) + \xi({\lambda-\lambda^k_\beta},{-\delta+\tfrac12}) \right]
\\
	&\quad
	+ \frac{1}{N}\sum^{k=2}_\beta \left[ \xi({\lambda-\lambda^k_\beta}, {\delta+\delta^k_\beta+2}) + \xi({\lambda-\lambda^k_\beta}, {-\delta-\delta^k_\beta}) 
	+\xi({\lambda-\lambda^k_\beta}, {\delta-\delta^k_\beta+1}) + \xi({\lambda-\lambda^k_\beta}, {-\delta+\delta^k_\beta+1})
	\right]
	\modulo{\pi}
\notag
\end{align}
where, for legibility, we dropped the indices $^{(2)}_\alpha$.
To compare with the Bethe--Takahashi equations, let us take all $\delta\goesto0$,
\begin{align}
	&\arctan{\lambda} + \frac{\pi}{2}  = 
	\pi\frac{J_+ + J_-}{N}
	+ \frac{1}{N}\sum^{k=1}_\beta \left[ \arctan\tfrac23(\lambda-\lambda^k_\beta) + \arctan 2({\lambda-\lambda^k_\beta}) \right]
\\
	&\hspace{4cm}
	+ \frac{1}{N}\sum^{k=2}_\beta \left[ \arctan\tfrac12({\lambda-\lambda^k_\beta}) + \frac\pi2\sign({\lambda-\lambda^k_\beta})
	+ 2\arctan({\lambda-\lambda^k_\beta}) 
	\right]
	\modulo{\pi}
\notag
\end{align}
Comparing to the Bethe--Takahashi equations, we conclude for the relation between the quantum numbers
\begin{align}
	I^{(2)} -  \half \sum_\beta^{k=2} \sign(\lambda^{(2)}-\lambda^{(2)}_\beta) = \step(\delta) + 2J_+ - \frac{N}{2}  \modulo {N}
\end{align}
taking this equation modulo $2$, we see that the criterion for a two-string with quantum number $I^{(2)}$ to be wide is
\begin{align}
\label{eq:wide_criterion}
 	\step(\delta) = I^{(2)} + \frac{N}{2} -M+1- \half\sum_\beta^{k=2}  \sign(\lambda^{(2)}-\lambda^{(2)}_\beta)  \modulo{2}
~.
\end{align}

The sum of the Bethe equations gives the equation for the string centers; therefore, we must look at the difference of the Bethe equations to find the deviations. 
\forget{
We have already seen in section \ref{sec:pairs} that the real part of this equation equals
\begin{align}
	J_n^+ - J_n^- &= -\step(\delta_n)
~;
\end{align}
}
The imaginary part of this equation can be written
\begin{align}
\label{eq:deviance}
	\left[\frac{1+\delta}{\delta}\right]^2 &= \left[\frac{(1+\delta)^2+\lambda^2}{\delta^2+\lambda^2} \right]^N 
		\prod^{k=1}_\beta \frac{(\delta-1/2)^2 + (\lambda-\lambda^k_\beta)^2}{(\delta+3/2)^2 + (\lambda-\lambda^k_\beta)^2}
		\prod^{k=2}_\beta \frac{(\delta+\delta^k_\beta)^2 + (\lambda-\lambda^k_\beta)^2}{(2+\delta+\delta^k_\beta)^2 + (\lambda-\lambda^k_\beta)^2}
			\frac{(1-\delta+\delta^k_\beta)^2 + (\lambda-\lambda^k_\beta)^2}{(1+\delta-\delta^k_\beta)^2 + (\lambda-\lambda^k_\beta)^2}
~.
\end{align}

Together, equations \eqref{eq:two_string_with_deviance}, \eqref{eq:deviance}, and \eqref{eq:wide_criterion}  determine the rapidity $\lambda^{(2)}_\alpha$ and deviation $\delta^{(2)}_\alpha$. It is of importance to note that the right-hand side of this equation is not strongly dependent on $\delta^{(2)}_\alpha$, as long as $\delta^{(2)}_\alpha$ is small compared to $\lambda^{(2)}_\alpha$, so that an iterative approach to solving these coupled equations can converge rapidly.


\subsection{Deviated three-strings}
Let us consider the equations for the deviation of a three-string in the presence of other strings of length not more than $2$, which is enough to illustrate the general idea.
In the presence of longer strings, terms will have to be added to these expressions but as these become quite unwieldy we defer this derivation to the treatment of the general case in the next section. 

Parametrising the three-string as 
\begin{align}
	\lambda^{(3)}_{\alpha\pm} &= \lambda^{(3)}_\alpha + \epsilon^{(3)}_\alpha \pm i(1+\delta^{(3)}_\alpha)
&
	\lambda^{(3)}_{\alpha0} &= \lambda ^{(3)}_\alpha
~,
\end{align}
we consider the sum over all three Bethe equations, writing $\lambda^{(3)}_{\alpha} = \lambda$
\begin{align}
\hspace{3cm}&\hspace{-3cm}
	\arctan 2\lambda + \xi({2\lambda+2\epsilon},{3+2\delta}) + \xi({2\lambda+2\epsilon}, {-1-2\delta}) 
	= \frac{\pi}{N}(J_- + J_0 + J_+)  \modulo{\pi}
\\
	+ \frac{1}{N} \sum_\beta^{k=1} 
		&\biggl[
			\arctan(\lambda-\lambda^k_\beta) 
			+ \xi({\lambda+\epsilon-\lambda^k_\beta},{2+\delta})+ \xi({\lambda+\epsilon-\lambda^k_\beta}, {-\delta})
		\biggr]
\notag\\
	+	\frac{1}{N} \sum_\beta^{k=2}  &\biggl[ 
			\xi({\lambda-\lambda^k_\beta}, {\tfrac32+\delta^k_\beta}) 
			+ \xi({\lambda-\lambda^k_\beta}, {\thalf-\delta^k_\beta})
			+ \xi({\lambda+\epsilon-\lambda^k_\beta}, {\tfrac52+\delta+\delta^k_\beta})
\notag\\
&\quad
			+ \xi({\lambda+\epsilon-\lambda^k_\beta}, {-\tfrac12-\delta-\delta^k_\beta}) 
			+ \xi({\lambda+\epsilon-\lambda^k_\beta}, {\tfrac32+\delta-\delta^k_\beta})
			+ \xi({\lambda+\epsilon-\lambda^k_\beta}, {\tfrac12-\delta+\delta^k_\beta}) 
\biggr]
\notag
~.
\end{align}
Taking its limit for $\delta \goesto 0, \epsilon\goesto 0$, 
\begin{align}
\hspace{5cm}&\hspace{-5cm}
	\arctan\frac{2\lambda}3 
	= \frac{\pi}{N}(J_- + J_0 + J_+)  
	+ \frac{1}{N} \sum_\beta^{k=1} 
		\biggl[
			\arctan(\lambda-\lambda^k_\beta) + \arctan\tfrac12({\lambda-\lambda^k_\beta})
			+ \frac\pi2\sign({\lambda-\lambda^k_\beta})
		\biggr]
\modulo{\pi}
\\
	+	\frac{1}{N} \sum_\beta^{k=2}  &\biggl[ 
			2\arctan\tfrac23({\lambda-\lambda^k_\beta}) 
			+ \arctan2({\lambda-\lambda^k_\beta})
			+ \arctan\tfrac25({\lambda-\lambda^k_\beta})
			+ \pi \sign({\lambda-\lambda^k_\beta})
\biggr]
\notag
~.
\end{align}

Comparing to the Bethe--Takahashi equations, we find
\begin{align}
	J_- + J_0 + J_+ &=  I^{(3)}_{\alpha} 
		-\half \sum_\beta^{k=1} \sign (\lambda -\lambda^k_\beta)
		- \sum_\beta^{k=2} \sign (\lambda -\lambda^k_\beta)
		\forget{-2\sum_\beta^{k=3} \sign (\lambda -\lambda^k_\beta)}
\modulo{N}
~.
\end{align}
The real part of the difference between the $+$ and $-$ logarithmic Bethe equations yields $J_+ - J_- = -1$. This leaves $J_0$ as of yet undetermined, but it turns out to be unnecessary to know this quantum number to be able to solve the equations.
 
The imaginary part of the difference between the $+$ and $-$ equations gives, when exponentiated,
\begin{align}
 	\delta^2+\epsilon^2  = r^2 &\definedby
		{[(2+\delta)^2 + \epsilon^2]} \left[\frac{3+2\delta}{1+2\delta}\right]^2 
		\left[ \frac{(1+2\delta)^2  + 4(\lambda+\epsilon)^2}{(3+2\delta)^2 +4(\lambda+\epsilon)^2} \right]^N 
\times
\notag\\
&\quad\times
		\prod_\beta^{k=1} \frac{(2+\delta)^2 +(\lambda+\epsilon-\lambda^k_\beta)^2}{\delta^2 +(\lambda+\epsilon-\lambda^k_\beta)^2} 
\,\times
\\
&\quad\times
		\prod_\beta^{k=2} \frac{(\frac52+\delta+\delta^k_\beta)^2 +(\lambda+\epsilon-\lambda^k_\beta)^2}{(\half+\delta+\delta^k_\beta)^2 +(\lambda+\epsilon-\lambda^k_\beta)^2} 
			\frac{(\frac32+\delta-\delta^k_\beta)^2 +(\lambda+\epsilon-\lambda^k_\beta)^2}
			{(\half-\delta+\delta^k_\beta)^2 +(\lambda+\epsilon-\lambda^k_\beta)^2} 
\notag
\forget{\\
&\quad\times
		\prod_\beta^{k=3} \frac{(3+\delta+\delta^k_\beta)^2 +(\lambda-\lambda^k_\beta+\epsilon-\epsilon^k_\beta)^2}{(1+\delta+\delta^k_\beta)^2 +(\lambda-\lambda^k_\beta+\epsilon-\epsilon^k_\beta)^2} 
			\frac{(1+\delta-\delta^k_\beta)^2 +(\lambda-\lambda^k_\beta+\epsilon-\epsilon^k_\beta)^2}{(1-\delta+\delta^k_\beta)^2 +(\lambda-\lambda^k_\beta+\epsilon-\epsilon^k_\beta)^2} 
			\frac{(2+\delta)^2 +(\lambda+\epsilon-\lambda^k_\beta)^2}{\delta^2 +(\lambda+\epsilon-\lambda^k_\beta)^2} 
\notag
}
\end{align}
The other independent equation is the Bethe equation for $\lambda^0$. However, for later generalisation it is more convenient to consider the sum of the $+$ and $-$ equations,
\begin{align}
\hspace{3cm}&\hspace{-3cm}
	\xi(\epsilon, -\delta)
	= \theta \definedby
	 - \xi({\epsilon}, {2+\delta})
	- \pi(J_++J_-) + N \left[\xi({2(\lambda+\epsilon)}, {3+2\delta}) + \xi({2(\lambda+\epsilon)}, {-1-2\delta}) \right]
\\
&\quad
	- \sum_\beta^{k=1} 
	\biggl[
	\xi({\lambda+\epsilon-\lambda^k_\beta}, {2+\delta}) + \xi({\lambda+\epsilon-\lambda^k_\beta},{-\delta})
	\biggr]
\notag\\
	&\quad
	- \sum_\beta^{k=2}	
	\biggl[
	\xi({\lambda+\epsilon-\lambda^k_\beta}, {\tfrac{5}{2}+\delta+\delta^k_\beta})
	+ \xi({\lambda+\epsilon-\lambda^k_\beta}, {-\tfrac{1}{2}-\delta-\delta^k_\beta} )
\notag\\
&\quad\quad\quad\quad
	+ \xi({\lambda+\epsilon-\lambda^k_\beta}, {\tfrac{3}{2}+\delta-\delta^k_\beta} )
	+ \xi({\lambda+\epsilon-\lambda^k_\beta}, {\tfrac{1}{2}-\delta+\delta^k_\beta} )
\biggr]
\notag
\end{align}
Both equations are written such that the terms on the right-hand side do not strongly depend on $\epsilon$ and $\delta$. 
Now we can simply iterate
\begin{align}
	\delta &= - \abs{r} \cos\theta
&
	\epsilon &= \abs{r} \sin\theta
~,
\end{align}
in order to find the exact roots.

\subsection{Deviated strings of any length}
Now that we know how to solve two- and three-strings, we are ready to generalise our approach to strings of any length.
Consider a generic $j$-string,
\begin{align}
	\lambda^j_{\alpha a} = \lambda^j_\alpha + \epsilon^j_{\alpha a} + \frac{i}{2}(j+1-2a) + i \delta^j_{\alpha a}
\end{align}
where the deviations $\delta$ and $\epsilon$ are real and satisfy $\delta_a = -\delta_{j+1-a}$, $\epsilon_a = \epsilon_{j+1-a}$. Furthermore, for even $j$, $\epsilon_{j/2} = 0$; for odd $j$, $\epsilon_{\floor{j/2}+1} = \delta_{\floor{j/2}+1} = 0$. 

Thus, for every $j$-string we have to find $\floor{j/2}$ deviations $d = \epsilon+i\delta$, as well as the string center $\lambda$. 
Following the logic of the three-string case, we will construct sets of equations for the norms and arguments of the 
deviations separately. These equations are written in such a way that the right-hand side can be used to calculate a 
new guess for the left-hand side, of which we then take the inverse function.  They are thus adapted to solution by
an iterative procedure.  For this to be successful, the equations are to be 
organised in such a way that the left-hand side varies strongly with the quantity under consideration (be it argument, norm, or 
center), whereas the right-hand side varies only weakly. Let us start with the arguments.

\subsubsection{Argument of deviations}
First we consider the sum of Bethe equations for two conjugate roots. 
The sum equation reads (again, indices $^j_\alpha$ are suppressed to reduce the strain on the eye)
\begin{align}	
\label{eq:sum_equation}
	\pi&(J_{a} + J_{j+1-a}) = \theta^a\sub{kin} - \theta^a\sub{other} - \theta^a\sub{self} \modulo{N\pi}
~,
\end{align}
where
\begin{align}
\label{eq:def_theta_deviation}
	\theta^a\sub{kin} &\definedby N \bigl[ \xi(2(\lambda+\epsilon_{a}), j+2-2a+2\delta_{a}) + \xi(2(\lambda+\epsilon_a), -j+2a-2\delta^j_{\alpha a}) \bigr] 
\\
	\theta^a\sub{other} &\definedby 
\hspace{-5mm}
	\sum_{k\beta}^{(k,\beta)\neq(j,\alpha)} 
\hspace{-5mm}
	\sum_{1\leq b \leq k}
		\xi\left(\lambda - \lambda^k_\beta + \epsilon_{a} - \epsilon^k_{\beta b}, 1+ (j-k)/2-(a-b) + (\delta_{a} -\delta^k_{\beta b})\right)
\notag\\
	&\hspace{17mm}
		+ \xi \left(\lambda - \lambda^k_\beta + \epsilon_{a} - \epsilon^k_{\beta b}, 1 - (j-k)/2 +(a-b) - (\delta_{a} -\delta^k_{\beta b})\right)
\notag\\
	\theta^a\sub{self} &\definedby \sum_{b=1}^{j} 
		\xi\left(\epsilon_{a} - \epsilon_{b}, 1-(a-b)+ (\delta_{a} - \delta_{b})
		\right)
 		+ \xi\left(\epsilon_{a} - \epsilon_{b}, 1+(a-b)- (\delta_{a} - \delta_{b})
		\right).
\notag
\end{align}

\forget{
Many of the sign terms in $\theta^a\sub{self}$ become independent of $\delta$ when $\abs\delta<1/2$. Singling out the terms for which this is not the case, 
we write the self-scattering term as
\begin{align}
	\theta^a\sub{self}&=\xi(
		{\epsilon_a - \epsilon_{a-1}},
		{\delta_a - \delta_{a-1}}
		)
	+ \xi(
		{\epsilon_{a} - \epsilon_{a+1}},
		-\delta_{a} + \delta_{a+1}
		)
\notag\\
	&\quad+ \arctan\frac
		{\epsilon_{a} - \epsilon_{a-1}}
		{2 -\delta_{a} + \delta_{a-1}}
	+ \arctan\frac
		{\epsilon_a - \epsilon_{a+1}}
		{2 +\delta_{a} - \delta_{a+1}}
\notag\\
	&\quad
	+
	\hspace{-5mm}
	\sum_{\substack{1\leq b \leq a-2 \\ a+2 \leq b \leq j-a-1}}
	\hspace{-5mm}
		\arctan\frac
			{\epsilon_{a} - \epsilon_{b}}
			{1-(a-b)+ (\delta_{a} - \delta_{b})}
 		+ \arctan\frac
			{\epsilon_{a} - \epsilon_{b}}
			{1+(a-b)- (\delta_{a} - \delta_{b})}
\notag\\
&\hspace{20mm}
		+\pi\sign(\epsilon_{a} - \epsilon_{b})
\notag
\end{align}
where it should be kept in mind that the terms which involve $a-1$ must be left out for $a=1$, as must terms with $a+1$ for $a=j$. 

Taking the equation modulo ${2\pi}$, we can now write
}
The self-scattering term  $\theta^a\sub{self}$ contains a few terms that vary strongly with $\epsilon, \delta$, if these are small. Let us split these off as
\begin{align}
	\theta^a\sub{self}&=  \theta^a\sub{residual} +
		\xi(
		{\epsilon_a - \epsilon_{a-1}},
		{\delta_a - \delta_{a-1}}
		)
	\ifwehave{a\neq1}
	+ \xi(
		{\epsilon_{a} - \epsilon_{a+1}},
		-\delta_{a} + \delta_{a+1}
		)
	\ifwehave{a\neq j}
~,
\end{align}
where the residual weakly-varying part is		
\begin{align}
\theta^a\sub{residual} &\definedby
 \xi(	
 		{\epsilon_{a} - \epsilon_{a-1}},
		{2 -\delta_{a} + \delta_{a-1}}
	)
	\ifwehave{a\neq1}
	+ \xi(
		{\epsilon_a - \epsilon_{a+1}},
		{2 +\delta_{a} - \delta_{a+1}}
	)
	\ifwehave{a\neq j}
\notag\\
	&\quad
	+
	\hspace{-5mm}
	\sum_{\substack{1\leq b \leq a-2 \\ a+2 \leq b \leq j-a-1}}
	\hspace{-5mm}
		\xi(
			{\epsilon_{a} - \epsilon_{b}},
			{1-a+b + \delta_{a} - \delta_{b}}
		)
 		+ \xi(
			{\epsilon_{a} - \epsilon_{b}},
			{1+a-b- \delta_{a} + \delta_{b}}
			)			
\end{align}
The strongly-varying terms are moved to the left-hand side in equation \eqref{eq:sum_equation},
\begin{align}
\label{eq:modified_sum_equation}
	\xi(\epsilon_{1} - \epsilon_{2}, -\delta_{1} + \delta_{2})
&= \theta_1 \modulo{2\pi}
\notag\\
 	\xi(\epsilon_a - \epsilon_{a-1}, \delta_a - \delta_{a-1})
	+ \xi(\epsilon_{a} - \epsilon_{a+1}, -\delta_{a} + \delta_{a+1})
&= \theta_a \modulo{2\pi} &&\text{for $1< a <j$}
\\
\xi(\epsilon_{j} - \epsilon_{j-1}, \delta_{j} - \delta_{j-1})
&= \theta_j \modulo{2\pi}
\notag
~,
\end{align}
where 
\begin{align}
\theta_a \definedby &- \pi(J_{a} + J_{j+1-a}) + \theta^a\sub{kin} - \theta^a\sub{other} -\theta^a\sub{residual}
~.
\end{align}
Using $\xi(\epsilon, \delta) + \xi(-\epsilon, \delta) = 0$, we can sum the above equations \eqref{eq:modified_sum_equation} to
\begin{align}
\label{eq:deviation_phase}
	\xi(\epsilon_{a} - \epsilon_{a+1}, -\delta_{a} + \delta_{a+1})
&=  \sum_{b=1}^a \theta_b \modulo{2\pi} 
~.
\end{align}
Applying the inverse function of $\xi$ on both sides, we determine $\epsilon_{a} - \epsilon_{a+1}$ and $\delta_{a} - \delta_{a+1}$ up to a common prefactor. 

Note that, to be able to use equation \eqref{eq:deviation_phase}, we need to determine $\pi(J_a + J_{j+1-a})$ modulo $2\pi$. Here we use
\begin{align}
	J_a + J_{j+1-a} &= 2J_a +1 = M \inlinemodulo{2} &&\text{if $a \neq j/2$}
\\
	J_{j/2} + J_{j/2+1} &= 2 J_{j/2} + (1 + \sigma)/2 = M - (1 + \sigma)/2\inlinemodulo{2} &&\text{if $a = j/2$}
~,
\end{align}
where, for even $j$, we need to use a criterion such as \eqref{eq:wide_criterion} to determine the inner pair sign, $\sigma\definedby \sign\delta_{j/2}$. We will defer this derivation to the end of this section.

\subsubsection{Norm of deviations}
To find the latter, we must consider the difference between the Bethe equations. Writing this as 
\begin{align}
	1 = r^2_{a,\text{kin}} \, r^{-2}_{a,\text{other}} \, r^{-2}_{a,\text{self}}
~,
\end{align}
with
\begin{align}
\label{eq:def_rsq}
	r^2_{a,\text{kin}} &\definedby 
		\left[ \frac
			{ (\lambda + \epsilon_{a})^2 + (j/2-a+\delta_{a})^2}
			{ (\lambda + \epsilon_{a})^2 + (j/2-a+1+\delta_{a})^2 }
		\right]^N
\notag\\
	r^2_{a,\text{other}} &\definedby 
		\hspace{-4mm}
		\prod_{k\beta}^{(k,\beta)\neq(j,\alpha)}
		\hspace{-5mm}
		\prod_{1\leq b \leq k}
		\hspace{-1mm}
			\left[ \frac
				{ (\lambda - \lambda^k_\beta + \epsilon_{a} - \epsilon^k_{\beta b})^2 
					+ (-1 + \frac{j-k}{2} -a+b 
					+\delta_{a}-\delta^k_{\beta b})^2 }
				{ (\lambda - \lambda^k_\beta + \epsilon_{a} - \epsilon^k_{\beta b})^2 
					+ (1 + \frac{j-k}{2} -a+b 
					+\delta_{a}-\delta^k_{\beta b})^2 }
			\right]
\notag\\
	r^2_{a,\text{self}} &\definedby 
		\prod_{1\leq b \leq j}^{b\neq a}
			\left[ \frac
				{ (\epsilon_{a} - \epsilon_{b})^2 
					+ (-1 -a+b 
					+\delta_{a}-\delta_{b})^2 }
				{ (\epsilon_{a} - \epsilon_{b})^2 
					+ (1  -a+b 
					+\delta_{a}-\delta_{b})^2 }
			\right]
~.
\end{align}
Again, we split up the self-scattering parts,
\begin{align}
	r^2_{a,\text{self}} &=  r^2_{a,\text{residual}}\,
	\frac{
	\left[
		{(\epsilon_a - \epsilon_{a+1})^2 + (\delta_a - \delta_{a+1})^2}
	\right]^{\ifwehave{a\neq j}}
	}
	{
	 \left[
		{(\epsilon_a - \epsilon_{a-1})^2 + (\delta_a - \delta_{a-1})^2}
	\right]^{\ifwehave{a\neq1}} 
	}
~,
\end{align}
where the residual part of the self-scattering term is
\begin{align}
r^2_{a,\text{residual}}\definedby
\frac
	{
	[(\epsilon_a - \epsilon_{a-1})^2 + (-2+ \delta_a - \delta_{a-1})^2]^{\ifwehave{a\neq1}}
	}	
	{
	[(\epsilon_a - \epsilon_{a+1})^2 + (2+\delta_a - \delta_{a+1})^2]^{\ifwehave{a\neq j}}
	}
	\prod_{1\leq b\leq j}^{b\not\in\set{a, a\pm 1}} 
		\frac
			{(\epsilon_a - \epsilon_b)^2 + (-1 - a+b + \delta_a - \delta_b)^2}
			{(\epsilon_a - \epsilon_b)^2 + (1 - a+b + \delta_a - \delta_b)^2}
\end{align}
Now we can write, in similar fashion as before,
\begin{align}
	{(\epsilon_1 - \epsilon_2)^2 + (\delta_1 - \delta_2)^2} &= r^2_1
\notag\\
\left[\frac
	{(\epsilon_a - \epsilon_{a+1})^2 + (\delta_a - \delta_{a+1})^2}
	{(\epsilon_a - \epsilon_{a-1})^2 + (\delta_a - \delta_{a-1})^2}
\right] &= r^2_a
\notag\\
\left[{(\epsilon_j - \epsilon_{j-1})^2 + (\delta_{j} - \delta_{j-1})^2}\right]^{-1} &= r^2_j
\notag
~,
\end{align}
and we may multiply these equations out to get the norm we sought,
\begin{align}
\label{eq:deviation_norm}
	{(\epsilon_a - \epsilon_{a+1})^2 + (\delta_a - \delta_{a+1})^2}
	&= \prod_{b=1}^a r^2_b
~.
\end{align}
In these equations,
\begin{align}
r^2_a &\definedby  r^2_{a,\text{kin}} \, r^{-2}_{a,\text{other}} \, r^{-2}_{a,\text{residual}}
~.
\end{align}
Equations \eqref{eq:deviation_norm} and \eqref{eq:deviation_phase} completely determine $\epsilon_a - \epsilon_{a+1}$ and $\delta_a - \delta_{a+1}$. For odd $j$, this sequence ends at $a=\floor{j/2}$, where $\epsilon_{a+1}=\delta_{a+1}=0$. For even $j$, the endpoint is at $j/2$, where $\epsilon_a=\epsilon_{a+1} = 0$ and $\delta_a=-\delta_{a+1}$. In both cases this allows us to find $\epsilon$ and $\delta$ themselves at the endpoint, after which all other deviations are found by summing the differences.

For odd $j$, therefore, the deviations are found from
\begin{align}
\label{eq:deviations_odd}
	\delta_a &= -\sum_{b=a}^{\floor{j/2}} \cos\Bigl[ \sum_{c=1}^b \theta_c\Bigr] \prod_{c=1}^b \abs{r_c}
\notag\\	
	\epsilon_a &= \sum_{b=a}^{\floor{j/2}} \sin\Bigl[ \sum_{c=1}^b \theta_c\Bigr] \prod_{c=1}^b \abs{r_c}
~.
\end{align}

For even $j$, the value $\theta_{j/2}$ cannot be determined as above; we must decide the width of the middle pair $\sigma\definedby\sign\delta_{j/2}$ on the basis of a criterion such as \eqref{eq:wide_criterion}, which we shall derive shortly. Here, the deviations are found from
\begin{align}
\label{eq:deviations_even}
	\delta_a &= 
		 \thalf\sigma \prod_{c=1}^{j/2} \abs{r_c} 
		-\sum_{b=a}^{j/2 -1} \cos\Bigl[ \sum_{c=1}^b \theta_c\Bigr] \prod_{c=1}^b \abs{r_c}
\notag\\	
	\epsilon_a &= \sum_{b=a}^{j/2 -1} \sin\Bigl[ \sum_{c=1}^b \theta_c\Bigr] \prod_{c=1}^b \abs{r_c}
~.
\end{align}

\subsubsection{Rapidities}
Finally, we need an equation for the rapidities. This is found from the total sum of the string Bethe equations, which becomes the Bethe--Takahashi equation for the string in the limit where the deviations vanish. Annoyingly, we need to pay attention to all branch cut terms to make the correct connection with the Bethe--Takahashi equation.
The self-scattering terms all cancel, so that the total sum reads
\begin{align}	
	\pi\sum_{a=1}^j J^j_{\alpha, a} &=  
	\sum_{a=1}^{j} 
	\theta^a\sub{kin} - \theta^a\sub{other} 
~.
\end{align}

The kinematic phases add up to
\begin{align}
	\sum_{a=1}^{j} \theta^a\sub{kin}
	&= N\Biggl[ \xi(2\lambda+ 2\epsilon_1, j+2\delta_1) 
	+\ifwehave{\text{$j$ even}} \xi(2\lambda, -2\delta_{j/2}) 
\\
	&~+ \sum_{a=1}^{\floor{(j-1)/2}} \xi(2\lambda+2\epsilon_{a+1}, j-2a+2\delta_{a+1}) + \xi(2\lambda+2\epsilon_{a}, -j+2a-2\delta_{a})
	\Biggr]
\notag
\end{align}

For the scattering phases, writing 
\begin{align}
	\xi^\pm_{ab} \definedby \xi(\lambda^j_\alpha - \lambda^k_\beta + \epsilon^j_{\alpha a} - \epsilon^k_{\beta b}, 1 \pm[ (j-k)/2 + b-a + \delta^j_{\alpha a} -\delta^k_{\beta b}])
\end{align}
and using
\begin{align}
	\sum_{b=1}^{k}
		 \xi^+_{ab} + \xi^-_{ab}
	&~=~ \xi^-_{a1} + \xi^-_{a2} + \xi^+_{a, k-1} + \xi^+_{ak} + \sum_{b=1}^{k-2}  \xi^+_{ab} + \xi^-_{a, b+2} 
~,
\end{align}
we can group terms together as 
\begin{align}
	\sum_{a=1}^{j} \theta^a\sub{other} 
	&=
\hspace{-5mm}
	\sum_{k\beta}^{(k,\beta)\neq(j,\alpha)} 
	\sum_{a=1}^{j} 
	\Bigl[\xi^-_{a1} + \xi^-_{a2} + \xi^+_{a, k-1} + \xi^+_{ak} + \sum_{b=1}^{k-2}  \xi^+_{ab} + \xi^-_{a, b+2}\Bigr]
~.
\end{align}

The iterative prescription for $\lambda\equiv\lambda^j_\alpha$ is then
\begin{align}
\label{eq:iterate_center}
	 &\xi(2\lambda+ 2\epsilon_1, j+2\delta_1) 
	+\ifwehave{\text{$j$ even}} \xi(2\lambda, -2\delta_{j/2})  
\\
&\quad=  
	\frac{\pi}{N}\sum_{a=1}^j J^j_{\alpha, a}
	+ \frac{1}{N} \sum_{a=1}^{j} \theta^a\sub{other} 
\notag\\
&\quad\quad
	- \sum_{a=1}^{\floor{(j-1)/2}} \xi(2\lambda+2\epsilon_{a+1}, j-2a+2\delta_{a+1}) + \xi(2\lambda+2\epsilon_{a}, -j+2a-2\delta_{a})
\notag
~.
\end{align}
These equations take the place of the Bethe--Takahashi equations when solving for a deviated string. However, to make the connection with those, we need to establish a relationship between the various quantum numbers used. This we can do by taking the limit as $\delta, \epsilon\goesto 0$. 

Using $\xi(\epsilon, \delta) + \xi(\epsilon, -\delta) = \pi\sign\epsilon$ and $\xi(\epsilon, 0) = (\pi/2)\sign\epsilon$, and taking the equation modulo $N\pi$, the kinetic phase goes to
\begin{align}
	\lim_{\substack{\epsilon\goesto 0 \\ \delta\goesto 0}}
	\sum_{a=1}^{j} \theta^a\sub{kin}
	&=  
		N\Bigl[ \arctan \frac{2\lambda}{j}
		+ \frac{j-1}{2}\pi \sign\lambda \Bigr] 
	\modulo {N\pi}
~.
\end{align}

For the scattering phase, we use
\begin{align}	
	\lim_{\substack{\epsilon\goesto 0 \\ \delta\goesto 0}} \sum_{a=1}^{j} \sum_{b=1}^{k-2}  \xi^+_{ab} + \xi^-_{a, b+2} = j(k-2)\pi \sign(\lambda^j_\alpha - \lambda^k_\beta)
	~
\end{align}
and
\begin{align}	
	 \lim_{\substack{\epsilon\goesto 0 \\ \delta\goesto 0}} 
	 	 ~&
	 	 \left[
	 	 \xi^-_{1,2} +  \xi^+_{1,k} 
	 	+ \xi^-_{j,1} + \xi^+_{j, k-1} +\sum_{a=1}^{j-1} \left(\xi^-_{a, 1} + \xi^-_{a+1, 2} + \xi^+_{a, k-1} + \xi^+_{a+1, k}\right)
	 	\right]
\notag\\
	&=	\xi(2(\lambda^j_\alpha - \lambda^k_\beta), \abs{j-k}) + \xi(2(\lambda^j_\alpha - \lambda^k_\beta), j+k) 
\notag\\
	&\hspace{10mm}+ (j-k)\step{(j - k)}\,\pi\sign(\lambda^j_\alpha - \lambda^k_\beta)
	+ 2
		\hspace{-4mm}
		\sum_{c=\abs{j-k}+2}^{j+k-2} 
		\hspace{-3mm}
		\xi(2(\lambda^j_\alpha - \lambda^k_\beta), c)
\notag\\
	&=	(1-\kron{j}{k})\arctan\frac{2(\lambda^j_\alpha - \lambda^k_\beta)}{\abs{k-j}} + \arctan\frac{2(\lambda^j_\alpha - \lambda^k_\beta)}{k+j}
\notag\\
	&\hspace{10mm}+ \left[(j-k)\step{(j - k)} + \thalf \kron{j}{k} \right]\,\pi\sign(\lambda^j_\alpha - \lambda^k_\beta)
\\
	&\hspace{10mm}+ 2
		\hspace{-4mm}
		\sum_{c=\abs{k-j}+2}^{k+j-2} 
		\hspace{-3mm}
		\arctan\frac{2(\lambda^j_\alpha - \lambda^k_\beta)}{c}
\notag
\end{align}
where the sum over $c$ is in steps of $2$ and 
the $\step$ term derives from the cancellation of terms between $k-j$ and $j-k$, which occurs if $j>k$. The last equality follows because all remaining terms have positive values for the second argument of $\xi$.

Observing that $j(k-2)/2 + (j-k)\step(j-k) = jk/2-\min(j, k)$, we may conclude that the relation between Bethe and Bethe--Takahashi quantum numbers is, modulo $N$,
\begin{align}
\label{eq:relation_between_quantum_numbers}
	\sum_{a=1}^{j} J^j_{\alpha, a} &= I^j_\alpha 
	+ N \frac{j-1}{2}
	-
\hspace{-5mm} 	 
	\sum_{k\beta}^{(k,\beta)\neq(j,\alpha)} 
\hspace{-5mm} 
	 \sign(\lambda^j_\alpha - \lambda^k_\beta) \left\{ \frac{jk}{2}-\min(j, k) + \thalf \kron{j}{k}\right\}
~.
\end{align}

\subsubsection{Width of innermost pair}
Given relation \eqref{eq:wide_narrow} between $J_a$ and $J_{j+1-a}$, we can use expression \eqref{eq:relation_between_quantum_numbers}  to determine the width of the innermost pair of an even string. Taking it modulo $2$, we find
\begin{align}
	\step(\delta_{j/2})
	&=  \half\Bigl[ 2I^j_\alpha + N(j-1) - j(M+1) -j-2
	-
\hspace{-5mm} 	 
	\sum_{k\beta}^{(k,\beta)\neq(j,\alpha)} 
\hspace{-5mm} 
	 \sign(\lambda^j_\alpha - \lambda^k_\beta) \left\{ jk - 2\min(j, k) + \kron{j}{k}\right\}
	 \Bigr]
 	\modulo{2}
~,
\end{align}
so that with $\sigma = 2\step(\delta_{j/2}) - 1$ we have found the last ingredient needed to find the deviations of even strings by equation \eqref{eq:deviations_even}.


\subsection{Summary of the method}
In the previous sections we have derived a number of equations for deviated strings, which can be solved by simple iteration. One starts from a solution of the Bethe--Takahashi equations \eqref{eq:log_bethe_takahashi}, giving initial locations for the string centers. Naturally, these solutions have all deviations zero, $\epsilon^j_{\alpha a} = \delta^j_{\alpha a} = 0$. These values will be the initial guess. Equations \eqref{eq:deviations_odd} or \eqref{eq:deviations_even} then give the next guess for the deviations $\epsilon^j_{\alpha a}$ and $\delta^j_{\alpha a}$, whereas equations \eqref{eq:iterate_center} are used to obtain a new guess for the string centers $\lambda^j_\alpha$. This procedure can be repeated until the desired level of convergence is achieved. Some results of this procedure are given in section \ref{appendix_finite}, where we present the complete solution of the Bethe Ansatz for the Heisenberg chains of $N=8$ and $N=10$ sites. Complete solutions up to $N=6$ can be found elsewhere in the literature (see e.g. [\onlinecite{KarbachMueller1998}]).




\section{Structure of string solutions}
\label{sec:structure}

Various authors have studied the fine structure of $2$-string solutions, starting with Bethe himself \cite{Bethe1931}, and followed by Vladimirov \cite{Vladimirov1984}
 and Essler, Korepin, and Schoutens \cite{Essler1992}. It is found that there are two branches of two-strings: narrow and wide ones. The wide strings lie on a curve in the complex plane such that, with increasing real part, the imaginary part of the roots diverges with the asymptote $\imag\lambda = \pm \real\lambda / \sqrt{N-1}$. The narrow strings get closer to the real line with increasing real part, and finally collapse onto it. This means that for high quantum numbers no narrow string solutions are available. Instead, extra solutions appear with two real roots.  Motivated by these results, we study the three- and four-string case. 

\subsection{Fine structure for three-stings}

\begin{figure}
\begin{center}
\includegraphics[width=8cm]{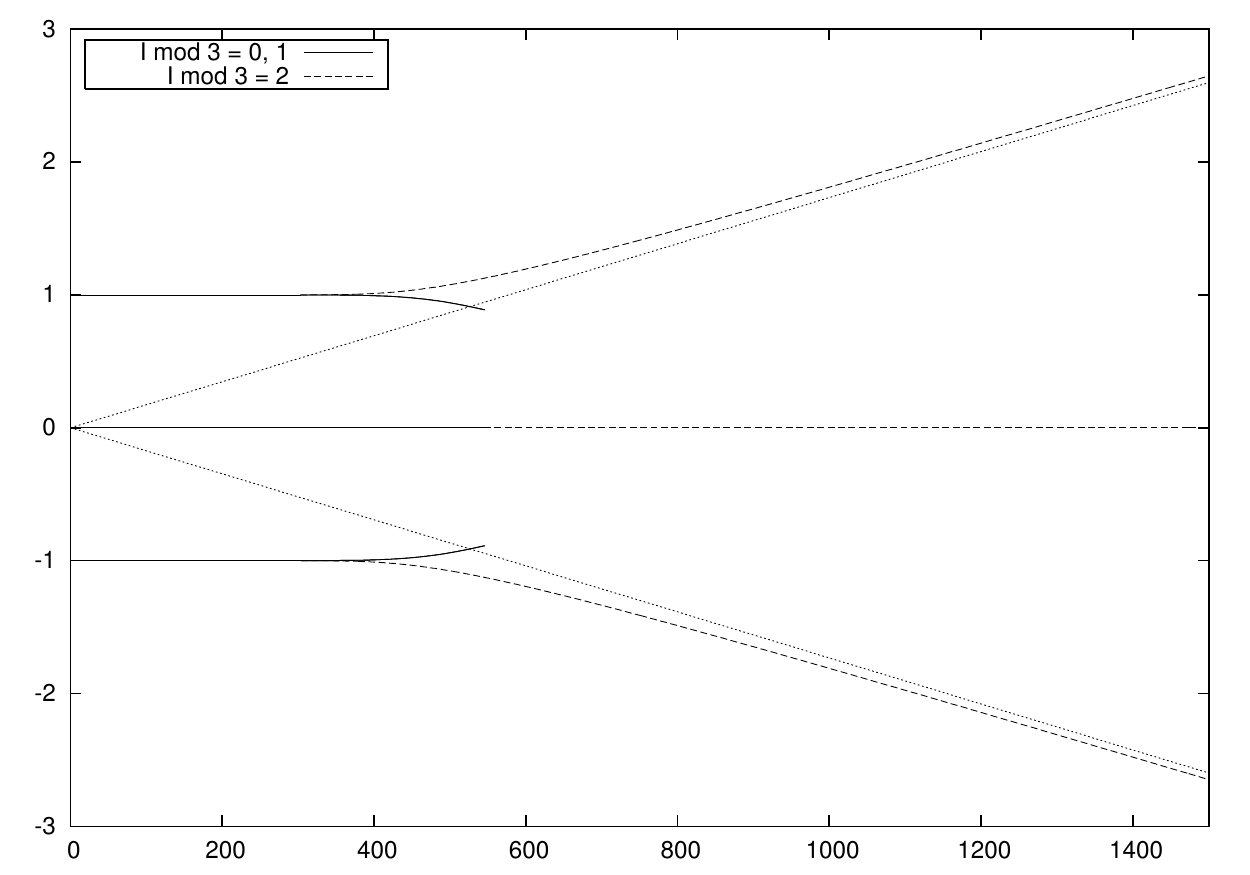}
\caption{Locus of three-strings for a chain with $N=10^6$ and $M=3$.  In this and all subsequent graphs, the rapidity complex plane is represented.  The straight lines are the asymptotes (\ref{eq:asymptote_3str}) for the branch with all three
quantum numbers distinct.}
\label{fig:threestring-all}
\end{center}
\end{figure}

Figure \ref{fig:threestring-all} shows the $\real\lambda >0$ solutions for a single three-string on a chain of $10^6$ sites. The solutions separate into three branches, distinguished by their quantum number $I \inlinemodulo{3}$. Two of the branches, with $I=0 \modulo{3}$ and $I=1 \modulo{3}$ are as good as indistinguisable for this value of $N$, although they are separately visible for smaller $N$.

\begin{figure}
\begin{center}
\includegraphics[width=8cm]{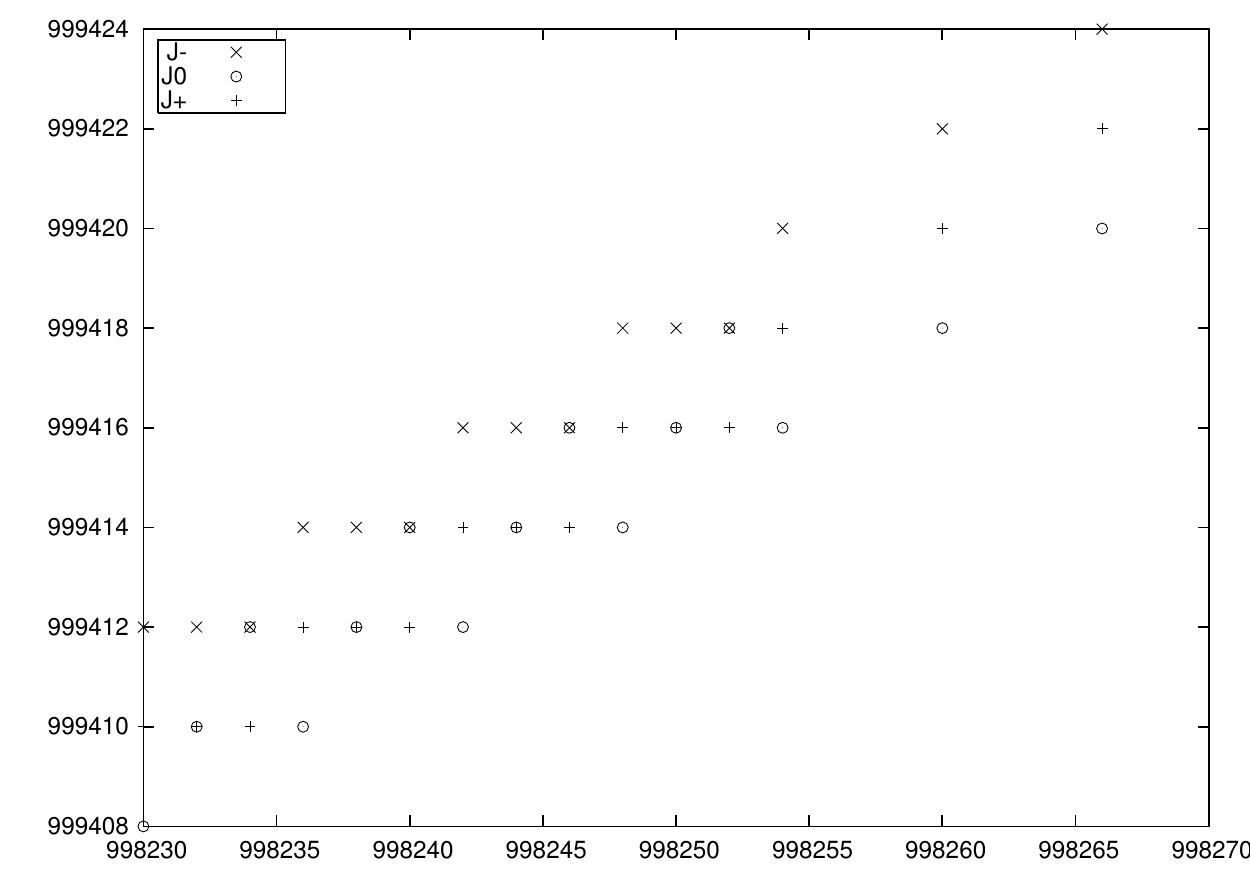}
\caption{Bethe quantum numbers $(J_-, J_0, J_+)$ for three-strings with increasing $I^{(3)}$. }
\label{fig:plot-3string-quantumnumbers}
\end{center}
\end{figure}

Calculating the Bethe quantum numbers $(J_+, J_0, J_-)$ from these solutions, it is found that they follow the pattern shown in figure \ref{fig:plot-3string-quantumnumbers}.
There are three branches of solutions, distinguished by the relations between the quantum numbers,
\begin{align}
	\text{branch $0$} && J_0+1 = &J_+ = J_--1 \notag\\
	\text{branch $1$} && J_0 = &J_+ = J_--1 \\
	\text{branch $2$} && J_0-1 = &J_+ = J_--1 \notag
\end{align}
Strings on the two branches for which $J_0 = J_+$ or $J_0 = J_-$ shrink with $\real\lambda$. 
Beyond a certain point we can no longer find a deviated three-string solution for the given quantum numbers. 
We conjecture that further states are made up of a narrow-string whose two roots have equal numbers, with the third quantum number
corresponding to the real root.  Further still, we expect the rapidities of states with yet higher quantum numbers to collapse onto
the real axis yielding a purely real three-string, similarly to what happens for two-strings. 
On the branch for which all quantum numbers are different, string solutions continue to exist with growing $\real\lambda$, with increasing deviation.  Assuming $\epsilon\ll\lambda$, we can derive the asymptotical relation
\begin{align}
	\imag \lambda &= \frac{\real\lambda}{\sqrt{\frac{1}{3}(N-3)}}
	~.
\label{eq:asymptote_3str}
\end{align}

\begin{figure}
\begin{center}
\includegraphics[width=8cm]{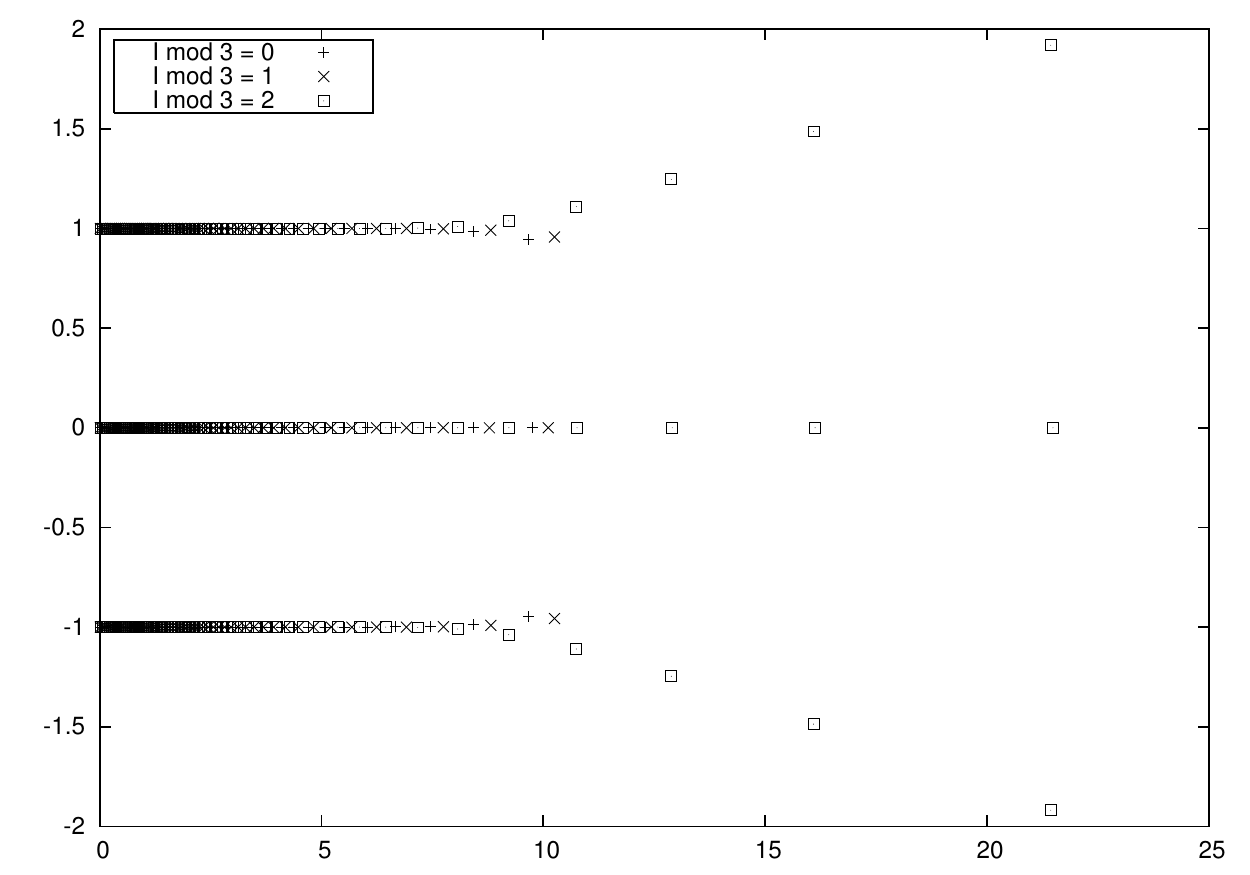}
\caption{Three-strings for a chain with $N=2000$ and $M/N=0.4$.}
\label{fig:s3_M0.4N}
\end{center}
\end{figure}

The results given up until now studied configurations where the string under consideration was the only set of roots. In the large-$N$ limit, this implies that we study a system very close to saturation field. To study a system at a magnetic field closer to zero, let us consider configurations with a single string accompanied by a number of real roots, such that $M/N=0.4$. This is numerically much more intensive as a far larger set of equations must be solved. Therefore, only smaller systems can be considered here. Figure \ref{fig:s3_M0.4N} shows the $\real\lambda >0$ solutions for a three-string in this case, on a chain of $2000$ sites. It turns out that the overall structure is very similar to the lone-string case considered earlier.

\subsection{Fine structure for four-strings}

\begin{figure}
\begin{center}
\includegraphics[width=8cm]{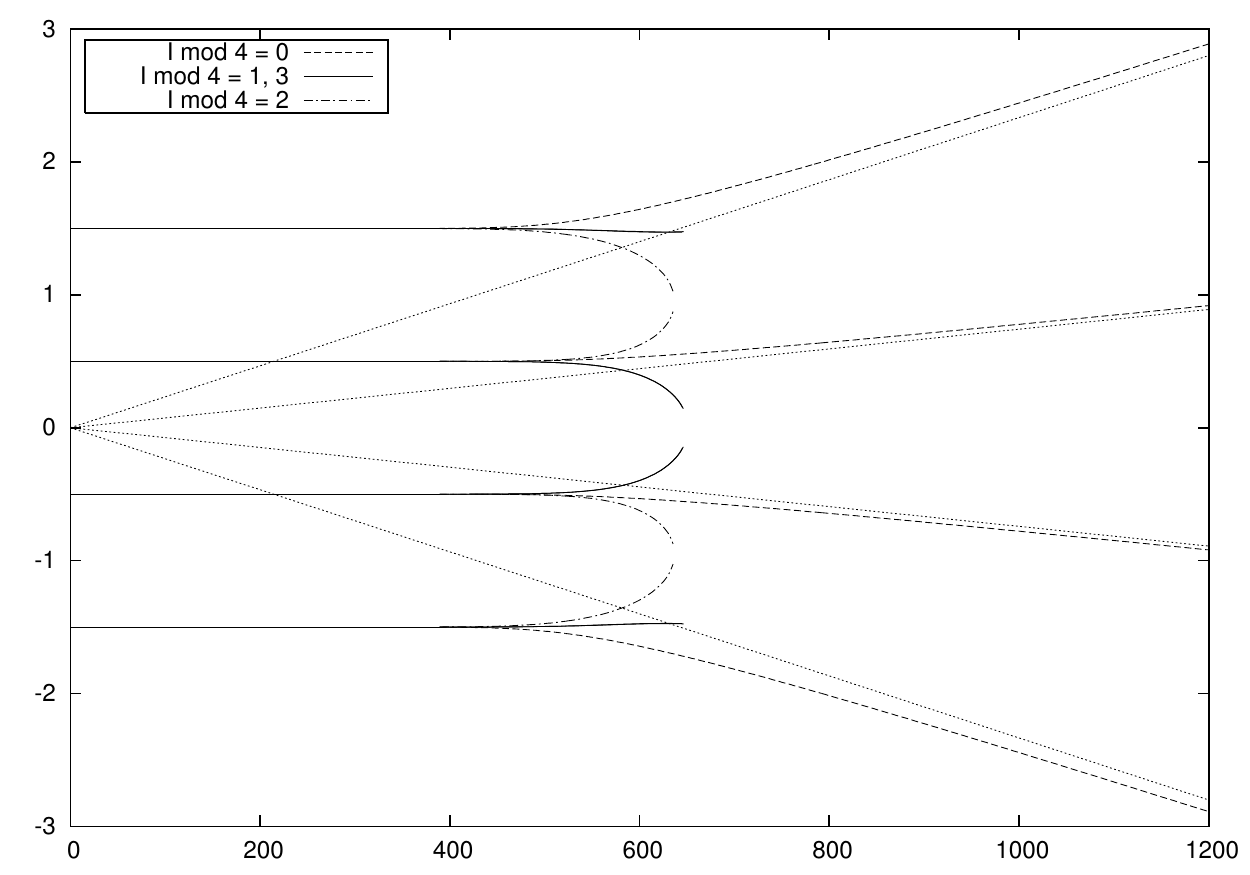}
\caption{Locus of four-strings for a chain with $N=10^6$ and $M=4$.  The straight lines are the asymptotes (\ref{eq:asymptote_4str}).}
\label{fig:fourstring-all}
\end{center}
\end{figure}

Figure \ref{fig:fourstring-all} shows the $\real\lambda >0$ solutions for a single four-string on a chain of $10^6$ sites. The solutions separate into four branches, distinguished by their quantum number $I \inlinemodulo{4}$. Again, two of these branches ($I = 1,3 \modulo{4}$) are nearly indistinguishable.

\begin{figure}
\begin{center}
\includegraphics[width=8cm]{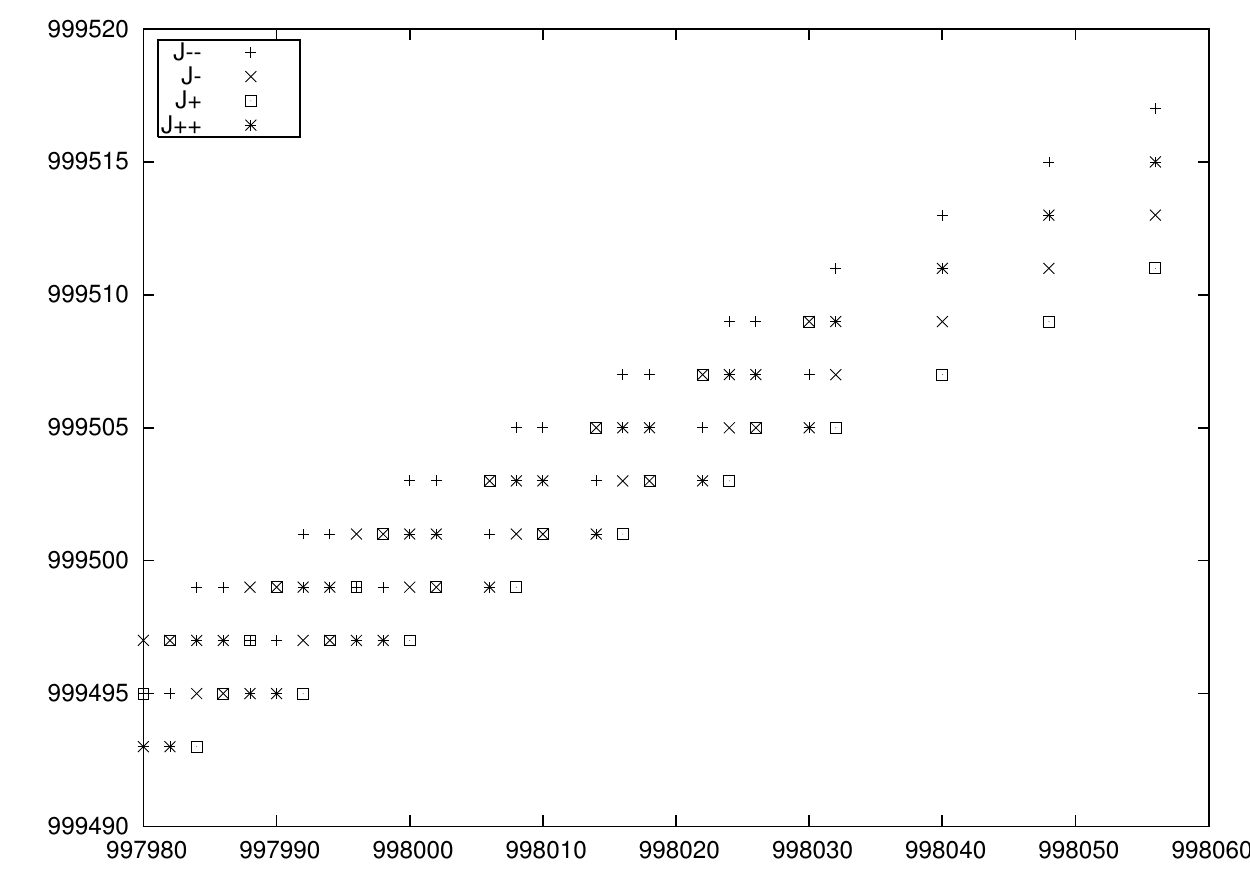}
\caption{Bethe quantum numbers $(J_{-2}, J_{-1}, J_{+1}, J_{+2})$ for four-strings with increasing $I^{(4)}$. }
\label{fig:plot-4string-quantumnumbers}
\end{center}
\end{figure}

Calculating the Bethe quantum numbers $(J_{-2}, J_{-1}, J_{+1}, J_{+2})$ from these solutions, it is found that they follow the pattern shown in figure \ref{fig:plot-4string-quantumnumbers}.
There are four branches of solutions, again distinguished by the relations between the quantum numbers,
\begin{align}
	\text{branch $0$} && J_{+1} &= J_{-1}-1 = J_{+2}-2 = J_{-2}-3 \notag\\
	\text{branch $1$} && J_{+1} &= J_{-1} = J_{+2}-1 = J_{-2}-2 \\
	\text{branch $2$} && J_{+1} &= J_{-1}-1 = J_{+2}+1 = J_{-2} \notag\\
	\text{branch $3$} && J_{+1} &= J_{-1} = J_{+2}+2 = J_{-2}+1  \notag
\end{align}

Assuming $\epsilon_0, \epsilon_1 \ll\lambda$, we can derive the asymptotical relation
\begin{align}
\label{eq:asymptote_4str}
	\imag \lambda &= \frac{\real\lambda}{\sqrt{N-3 \pm \sqrt{\frac{2}{3}(N-3)(N-2)} }}
\end{align}

\begin{figure}
\begin{center}
\includegraphics[width=8cm]{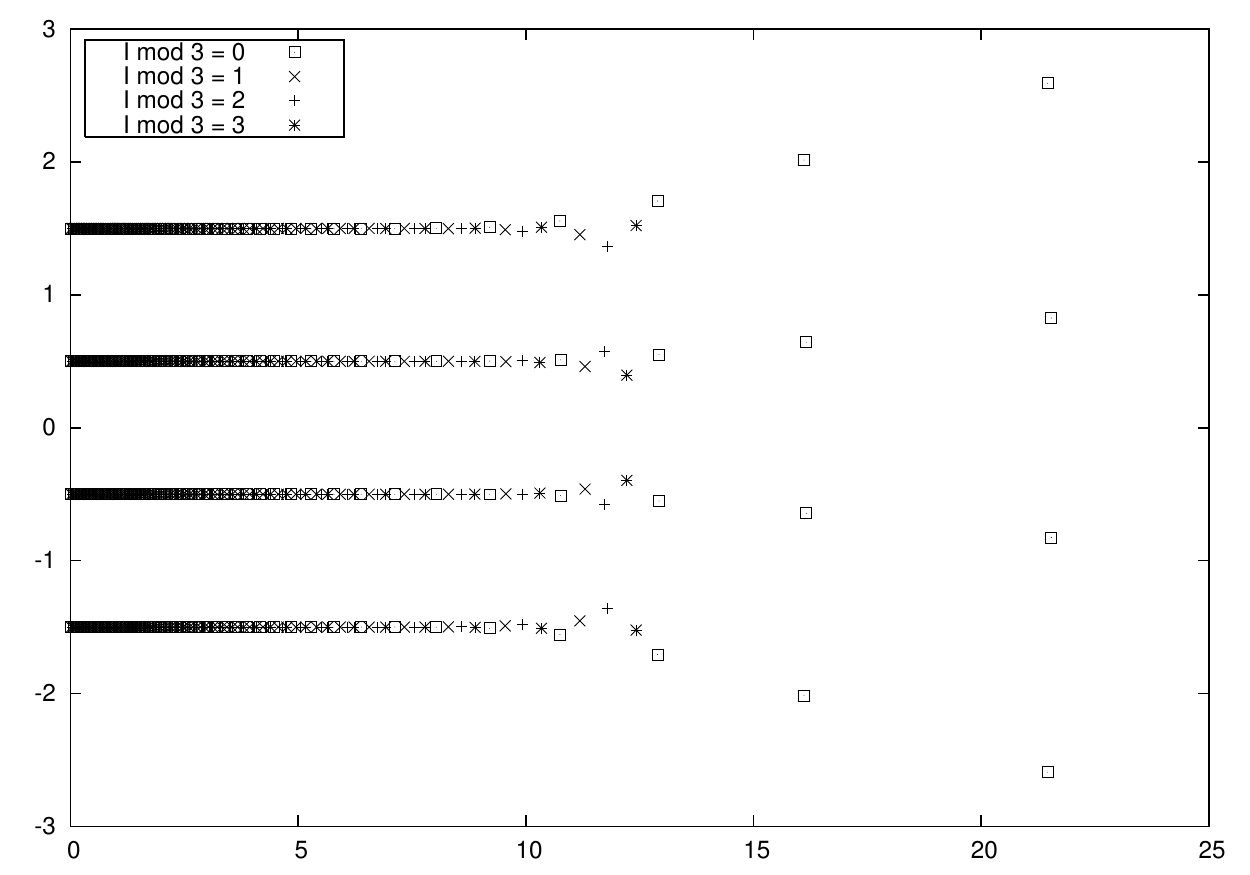}
\caption{Four-strings for a chain with $N=2000$ and $M/N=0.4$.}
\label{fig:s4_M0.4N}
\end{center}
\end{figure}

Figure \ref{fig:s4_M0.4N} shows the $\real\lambda >0$ solutions for a four-string accompanied by real roots such that $M/N=0.4$, on a chain of $2000$ sites. 


\subsection{Completeness}
An interesting question is the fate of string-like solutions with increasing chain length $N$. The analytic solution is known only for the two-magnon sector $M=2$; from $N=22$ onward narrow two-strings `collapse' and form pairs of roots on the real line, as described 
in [\onlinecite{Essler1992}]. The number of missing solutions equals
\begin{align}
\label{eq:missing-M2}
	n\sub{missing}^{(M=2)} = \floor{\frac{\sqrt{N}}{\pi} -\half} ~.
\end{align}

\begin{figure}
\begin{center}
\includegraphics[width=6cm]{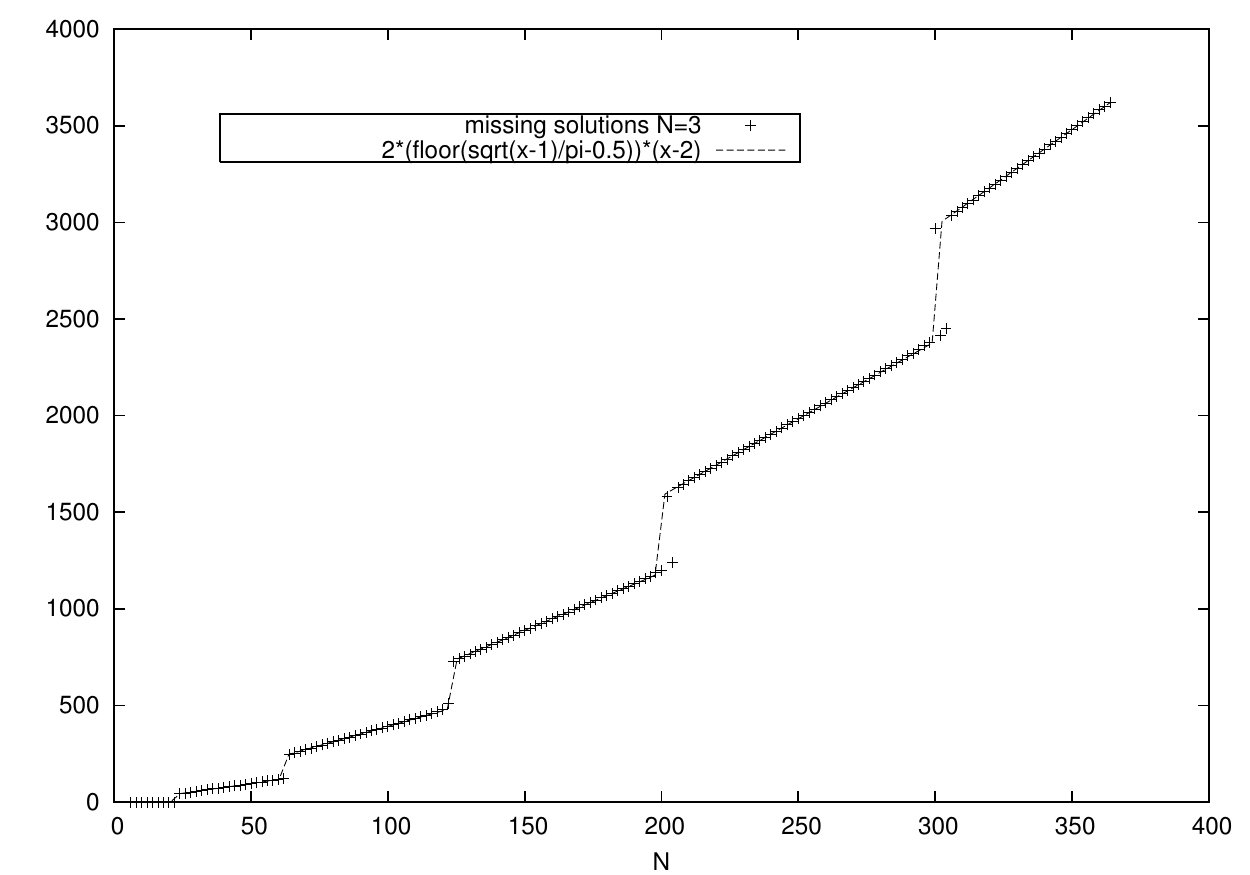}
\caption{Missing solutions at $M=3$.  Around the jumps it sporadically happens that solutions are not found, or wrong solutions are found, depending on the convergence threshold set. 
The number of missing solutions turns out to fit to an empirical rule much like equation \eqref{eq:missing-M2}, namely $n\sub{missing}^{(M=3)} = 2(N-2)\floor{\frac{\sqrt{N-1}}{\pi}-\half}$.}
\label{fig:missing-M3}
\end{center}
\end{figure}

\begin{figure}
\begin{center}
\includegraphics[width=6cm]{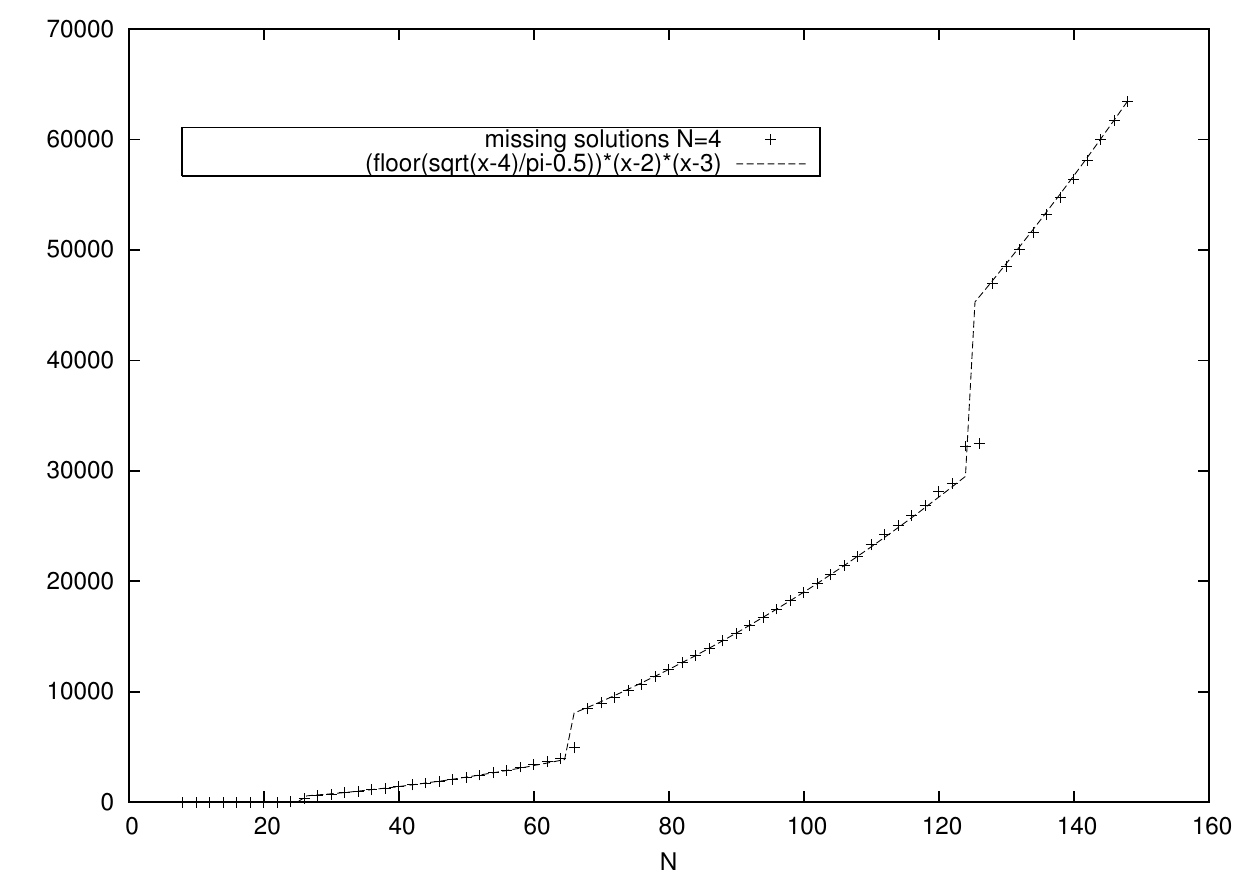}
\caption{Missing solutions at $M=4$. An empirical fit for the number of missing solutions is $n\sub{missing}^{(M=4)} = (N-2)(N-3)\floor{\frac{\sqrt{N-4}}{\pi}-\half}$.}
\label{fig:missing-M4}
\end{center}
\end{figure}

The method described in this chapter can shed some light on this question, as we can try and find all solutions of the Bethe equations for a given number of magnons with increasing chain length. The number of missing solutions then gives an upper bound to the number of non-string solutions. For $M=3$, this is shown in figure \ref{fig:missing-M3}. 
The situation for $M=4$ is shown in figure \ref{fig:missing-M4}.
Note that, especially around the jumps in the graphs, the number of solutions found is rather sensitive to the degree of convergency required in iteration. This effect may shift the jumps a bit to the left and right but the overall shape of the function is not changed.
	
\begin{figure}
\begin{center}
\includegraphics[width=6cm]{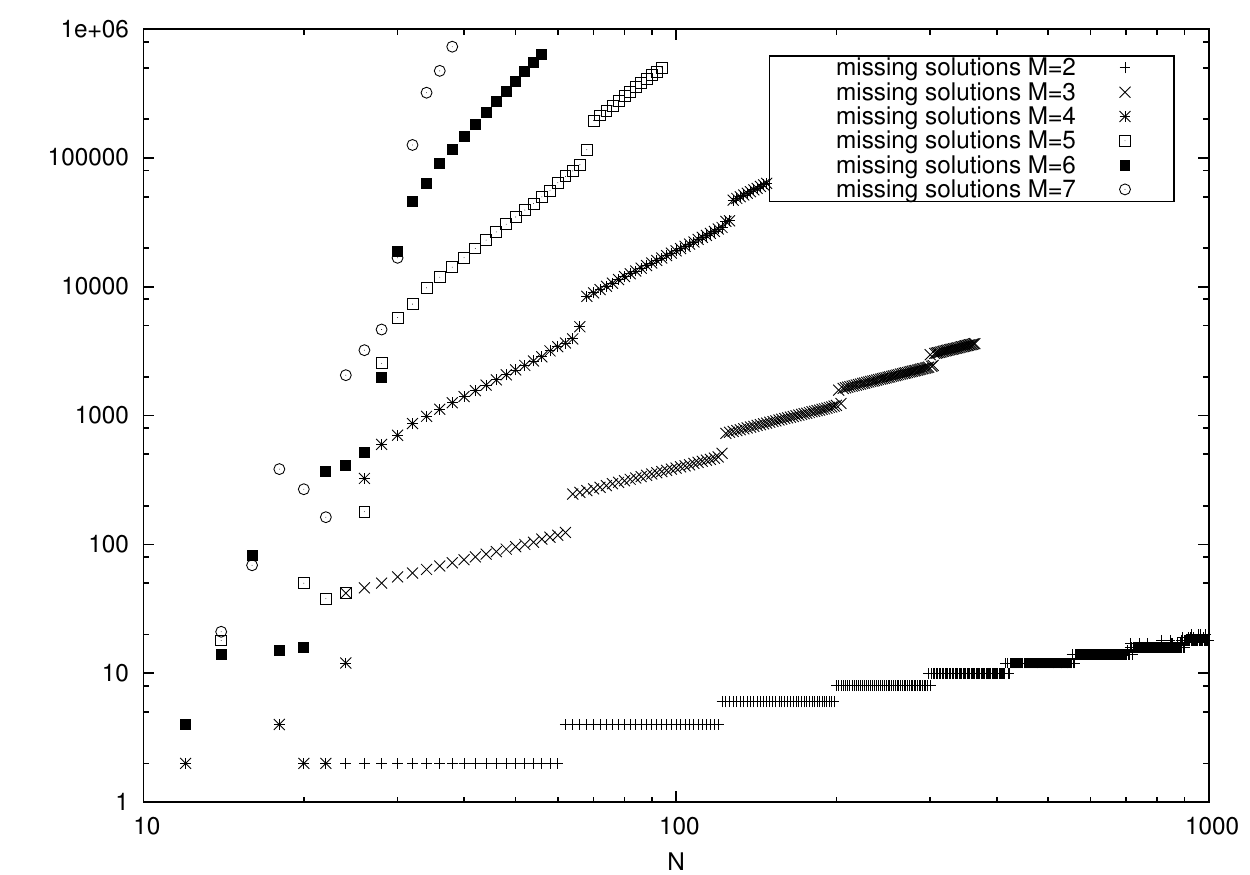}
\caption{Log-log plot of number of missing solutions at $M\in\set{2\ldots 7}$.}
\label{fig:missing}
\end{center}
\end{figure}

Finally, in figure \ref{fig:missing} we show the number of missing solutions in a single log-log graph for $M$ between $2$ and $7$. It can be seen from the plot  that at every $M$, the number of missing solutions grows as $O(N^{M-3/2})$ and exhibits jumps on or very close to the locations dictated by the $M=2$ rule \eqref{eq:missing-M2}. These numbers fit in a picture where the collapse of narrow pairs---either only from two-strings or from higher strings as well---is the only aberration from the string hypothesis, if one allows for deviations in the strings themselves. 
Of course, it would be desirable to have a method to solve for the collapsed pairs as well, so that this statement can be checked. Since the number of higher strings is much lower than that of two-strings, a collapse of higher strings would not make a big difference in these graphs.

This also shows that, at least at the small numbers of magnons and chain lengths considered here, the method we describe captures the vast majority of solutions of the Bethe equations: the number of highest-weight solutions is $n\sub{total}^{(M)}=\binom{N}{M} - \binom{N}{M-1} =O\left(\frac{N^M}{M!}\right)$ so that the fraction of missing solutions scales as
	$\frac{n\sub{missing}^{(M)}}{n\sub{total}^{(M)}} = \bigoh{\frac{N^{-3/2}}{M!}}$
	~.


\subsection{Scaling of deviations}
\begin{figure}
\begin{center}
\includegraphics[width=6cm]{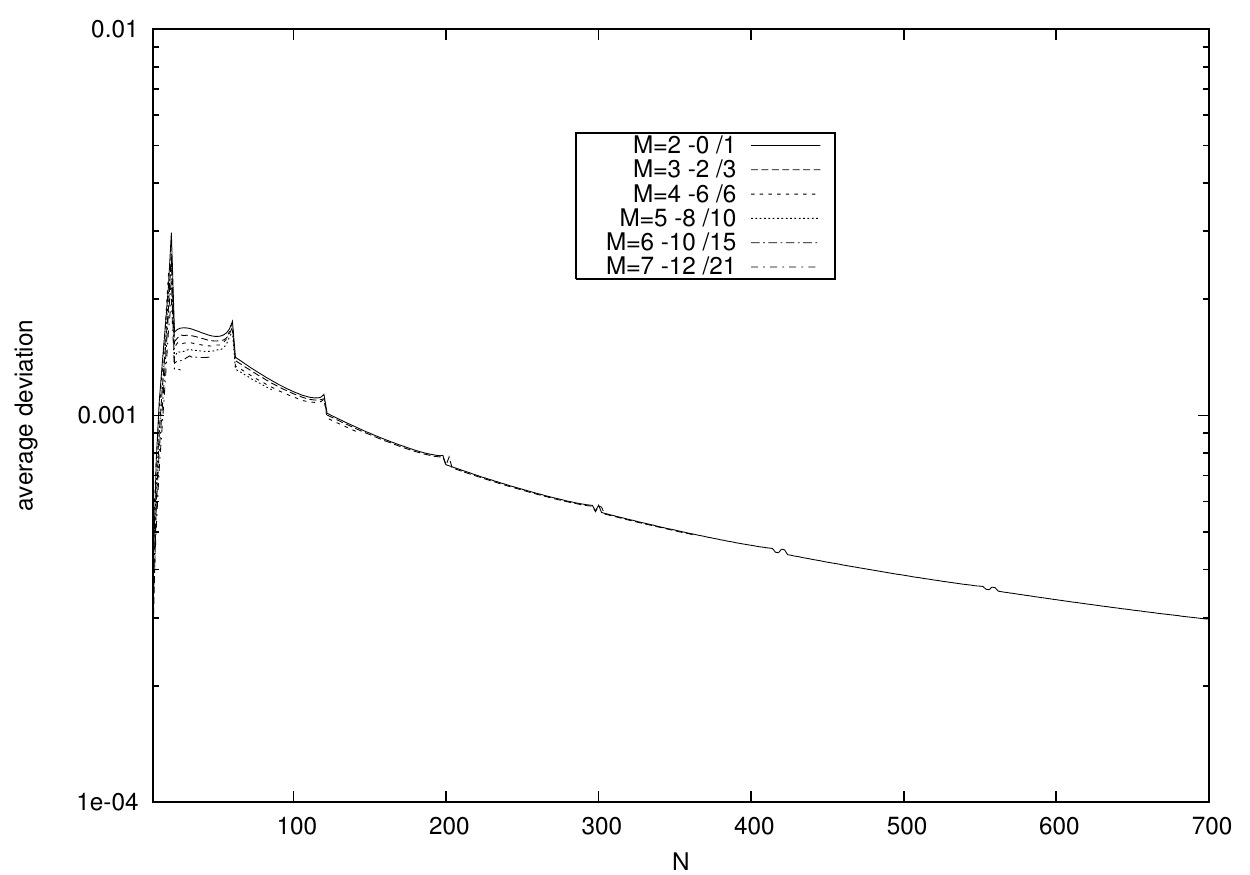}
\caption{Scaling of deviations with increasing $N$, plotted against $N-2M$. The vertical axis is logarithmic. The curves for $M>2$ are divided by $3(M-2)$ to show the collapse.}
\label{fig:deviation_scaling}
\end{center}
\end{figure}

The string hypothesis states that deviations should decrease with increasing chain length as $O(e^{-c N})$ for some constant $c$. In figure \ref{fig:deviation_scaling} the average deviation of string solutions is shown, where the average is taken over all solutions (including real roots) and the magnitude of deviation for a single string is given by
\begin{align}
	d \definedby \sum_{a=1}^{n_j} \delta_a^2 + \epsilon_a^2
~.
\end{align}
Remarkably, the curves for all values of $M$ collapse onto that of $M=2$, if the average deviations are divided by $\sum_{l=1}^{n_j-1} l$. The string hypothesis suggests that such an average should decrease as a sum of exponentials; the curve is consistent with this. 

\begin{figure}
\begin{center}
\includegraphics[width=8cm]{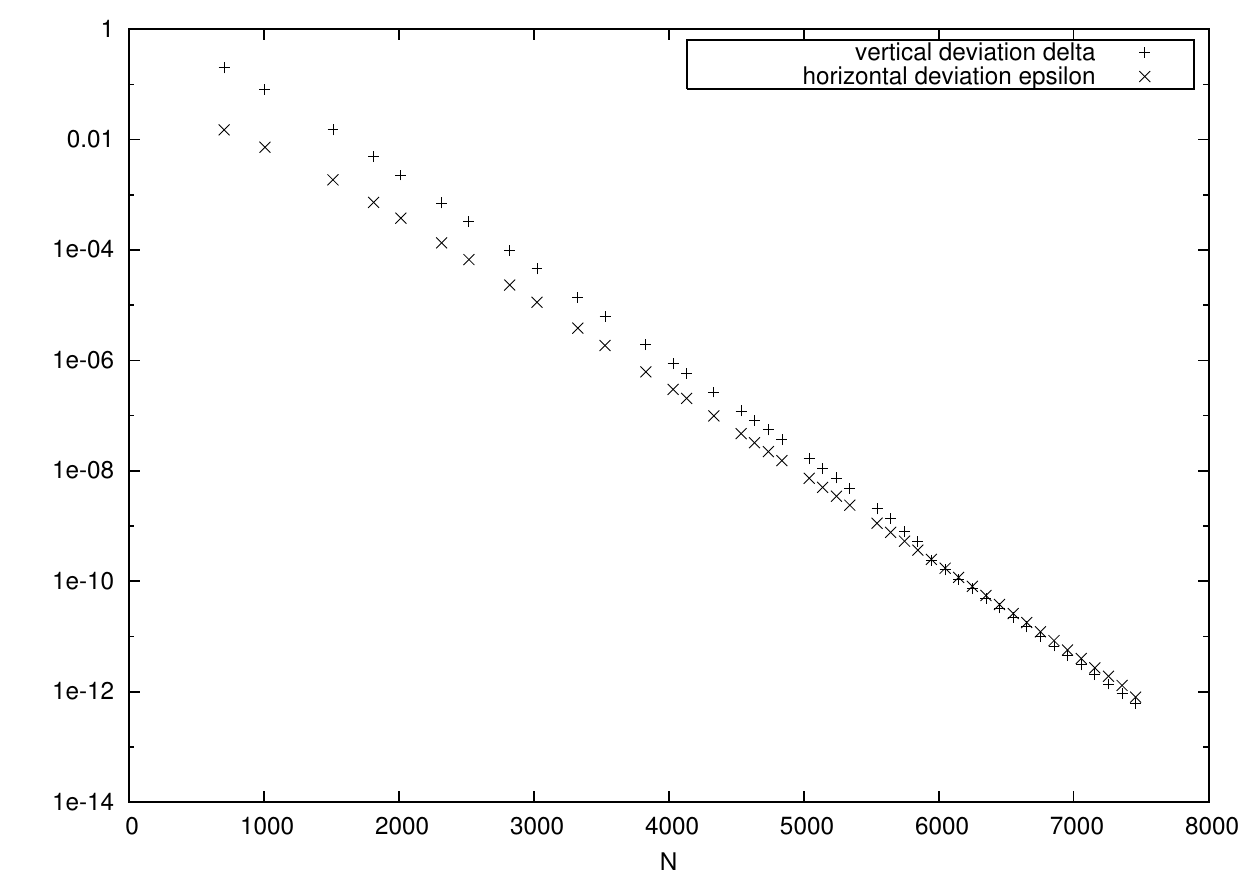}
\caption{Exponential decrease with chain length $N$ of the $\delta$ and $\epsilon$ of three-string solutions with $15.9925<\real\lambda<15.9975$ at $M=3$.}
\label{fig:plot-s3_fixed-lambda}
\end{center}
\end{figure}

We can also check the string hypothesis directly:
if we keep $\lambda$ approximately fixed (of course we are limited in our choice for $\lambda$, 
bounded as we are by having to find an actual solution of the Bethe equations) we see that the deviation $\delta$ indeed decreases exponentially with increasing $N$. This is shown in figure \ref{fig:plot-s3_fixed-lambda}.

\begin{figure}
\begin{center}
\includegraphics[width=8cm]{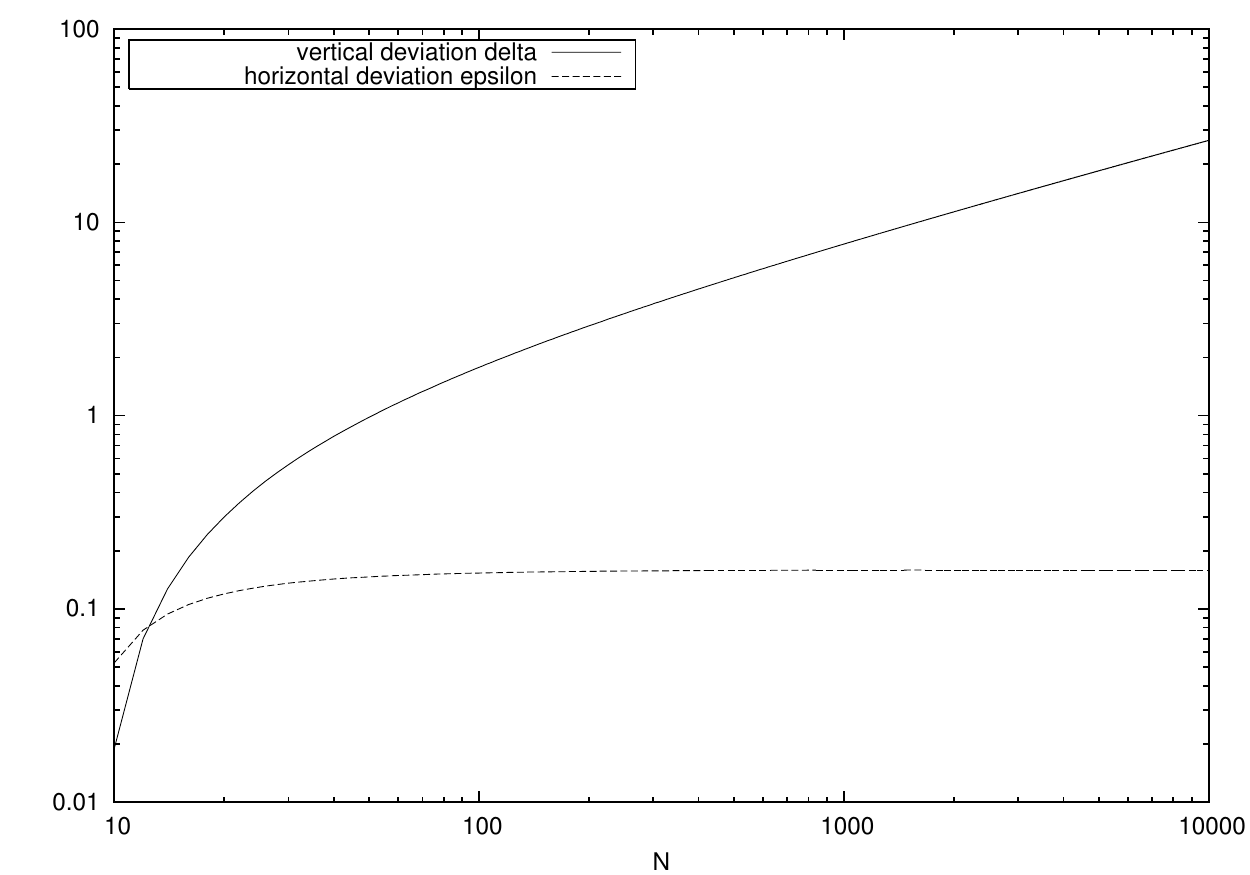}
\caption{Increase with chain length $N$ of the $\delta$ and $\epsilon$ of peripheral three-string solutions. The horizontal deviation $\epsilon$, which appears to saturate, in fact keeps growing.}
\label{fig:plot-s3_max-string}
\end{center}
\end{figure}

However, if we do not hold the string center constant but instead consider the behaviour of the most outward string, we see that its deviation in fact increases with $N$, as shown in figure \ref{fig:plot-s3_max-string}. This is because the string center of the peripheral string increases with $N$ as well. 


\subsection{Deviation of peripheric strings}

Let us parametrise the deviations as follows. For the roots in an $j$-string,
\begin{align}
	\lambda_a &= \lambda + \frac{i}{2}\left(j +1 -2a\right) + d_a 
\end{align}
where $d_a \definedby i\delta_a + \epsilon_a$. For the central root or pair,
\begin{align}
	\delta_{(j+1)/2} = \epsilon_{(j+1)/2} &= 0 &	&\text{ for $j$ odd}
\notag\\
	\epsilon_{j/2} &= 0 &	&\text{ for $j$ even}
\end{align}
Furthermore, as the roots are grouped in pairs of conjugates,
\begin{align}
	d_{j-a} = \conj{d_a}
\end{align}

In the Bethe equation for a given root, the scattering phase  features a product over the other roots of the string,
\begin{align}
	R_a &\definedby \prod_b^{b\neq a} 
	\frac	{ d_a - d_b + (b-a+1)i	}
			{ d_a - d_b + (b-a-1)i 	}
\end{align}

To lowest order in $d$, for $a\neq 1,j$,
\begin{align}
	R_a &\approx 
	- \left[\frac{d_a - d_{a-1} }{d_a-d_{a+1} }\right]
	\frac	{\prod_b^{b\neq a-1} (b-a+1)i }
			{\prod_b^{b\neq a+1} (b-a-1)i}
	=
	\left[\frac{d_{a-1} - d_a}{d_a-d_{a+1} }\right]
	\frac	{ (j-a+1) (j-a) }
			{ a (a-1)	}
~.
\end{align}

For $a=1$ or $a=j$,
\begin{align}
	R_1 &\approx 
	-i\frac{j (j-1) }{d_1-d_{2}}
	&
	R_j &\approx 
	-i\frac{ d_j - d_{j-1} }{ -j (1-j)}
\end{align}
Thus, starting from $a=1$, we can successively construct the differences 
\begin{align}
 	d_{a+1}-d_a &=  -i\frac{j(j-a)}{a}  \left[ \prod_{b=1}^{a} K_b \right] \left[\prod_{b=1}^{a-1}\frac{j-b}{b} \right]^2
 ~. 
\end{align}
where $K_a$ is given by the other factors in the Bethe equation,
\begin{align}
	K_a &\definedby
	\left[\frac{ \lambda + (j+2-2a)i/2 }{ \lambda + (j-2a)i/2  } \right]^{-N}
	\prod_{k\beta b}^{(k\beta)\neq(j\alpha)} \left[ \frac{ \lambda - \lambda^k_{\beta b} + (j+3-2a)i/2 }{  \lambda - \lambda^k_{\beta b} + (j-1-2a)i/2 } \right]
~;
\end{align}
the product is understood to run over all complex roots {\em not} belonging to the same string.

For odd $j$, we can use $d_{(j+1)/2} = 0$ and sum over the differences to get
\begin{align}
	d_b &= -\sum_{a=b}^{(j-1)/2}  \frac{j (j-a)}{a} 
	{\binom{j-1}{a-1}}^2
	\prod_{c=1}^{a} K_c 
	& &\text{for }  b \leq (j-1)/2
~,
\end{align}

For even $j$, we have $\real d_{j/2} = 0$ and therefore $d_{j/2} = - d_{j/2+1}$. We now have, $\text{for }  b \leq j/2-1$,
\begin{align}
	d_b &=  
		\frac{i}{2}\left[ j{\binom{j-1}{j/2-1}}^2 \prod_{c=1}^{j/2} K_c \right] 
		-
		i\sum_{a=b}^{j/2-1} \left[ \frac{j(j-a)}{a} {\binom{j-1}{a-1}}^2 \prod_{c=1}^{a} K_c \right] 
\end{align}

The behaviour of $K_a$ for large $N$ and $\lambda$ depends on the order in which we take the limits.
In particular, let us consider the limit for large $\lambda$, i.e. a string center far removed from the origin, while the other rapidities remain small.
Then, $\lim_{\lambda\goesto\infty} K_a  = 1$ and, for odd $j$,
\begin{align}
	\lim\limits_{\lambda\goesto\infty} 
	d_b &= -i\sum_{a=b}^{(j-1)/2} \frac{j(j-a)}{a} \binom{j-1}{a-1}^2
\end{align}
Note that this number is  of order unity and  independent of the chain length $N$. This order of limits is relevant for instance when we consider the limit to large $N$ at a fixed magnetisation density $M/N$, as in that case the number of rapidities grows with $N$. Then, assuming a constant rapidity density, strings on the periphery will always be strongly deviated.




\section{Symmetric and singular states}
\label{sec:central_strings}
\label{sec:symmetric}

If the Bethe--Takahashi quantum numbers are distributed symmetrically around zero, then so are the rapidities. Such symmetric states merit special attention. The simplest example is the ground state, which has already been discussed. Since all the rapidities are then real, no problem is encountered. In the presence of bound states, the situation becomes more complicated.  In a symmetric state, if the quantum number of a string is zero, then its center is also
zero.  Superimposing two higher strings of length differing by an even integer ({\it e.g.} a 2-string and a 4-string, or a 3-string and a 7-string)
means that pairs of rapidities coincide in the pure string hypothesis.  String deformations, as we will see, regularize these situations and
give allowable eigenstates.  As we will discuss in this section, it turns out to distinguish two classes of symmetric states: 
those with only strings, and those which include even strings. 
Again, we concentrate on the isotropic chain, although similar issues exist also in the XXZ chain.

\subsection{Multiple symmetric odd strings}
In a symmetric state with more than one odd string at the origin, the solution of the Bethe--Takahashi equations is not a valid state: since there are two or more roots present at the origin, the exclusion principle is violated. However, if we take the deviations into account, this problem does not arise: the two roots that coincide in the limit are actually separated.
As an example, we will show the solution for the simplest case where this problem arises: the symmetric state with one three-string and one real root at the origin, $I^{(3)}=0$ and $I^{(1)}=0$. Defining the rapidities $\lambda_{\pm i} = \pm(i+\delta)$, $\lambda_{-0} = -\lambda_{+0} >0$, we find from the difference of Bethe equations that $J_{+i} = -J_{-i} =(N-1)/2$ (where also $J_{-i} = J_{+i} +1 \inlinemodulo{N}$). The quantum numbers for the real roots in the complex must be half-integer, opposite, and as small as possible, leading to $J_{\pm0} = \pm1/2$. 

The fixed points of the iterative equations for the deviations as given in section \ref{sec:deviations}, however, are repulsive in this case; therefore we need to either use another method (such as Newton--Raphson) or rewrite the iterative equations. An easy prescription that works is to take the sum of the equations for the positive-real and positive-imaginary root, which gives
\begin{align}
\label{eq:odd_sym_iter}
	\lambda' + i(1+\delta') = \tan\left[\frac{N-1}{2}\bigl(\arctan 2\lambda + i \arctanh(2+2\delta)\bigr) -\frac{\theta\sub{other}}{2}\right]
~,
\end{align}
where, in the presence of other roots,
\begin{align}
\theta\sub{other} \definedby \sum_\beta^{\lambda_\beta\not\in\set{\pm\lambda, \pm i(1+\delta)}}
	\arctan (\lambda -\lambda_\beta) + \arctan(i+i\delta -\lambda_\beta)
~.
\end{align}

\begin{figure}
\begin{center}
\includegraphics[width=6cm]{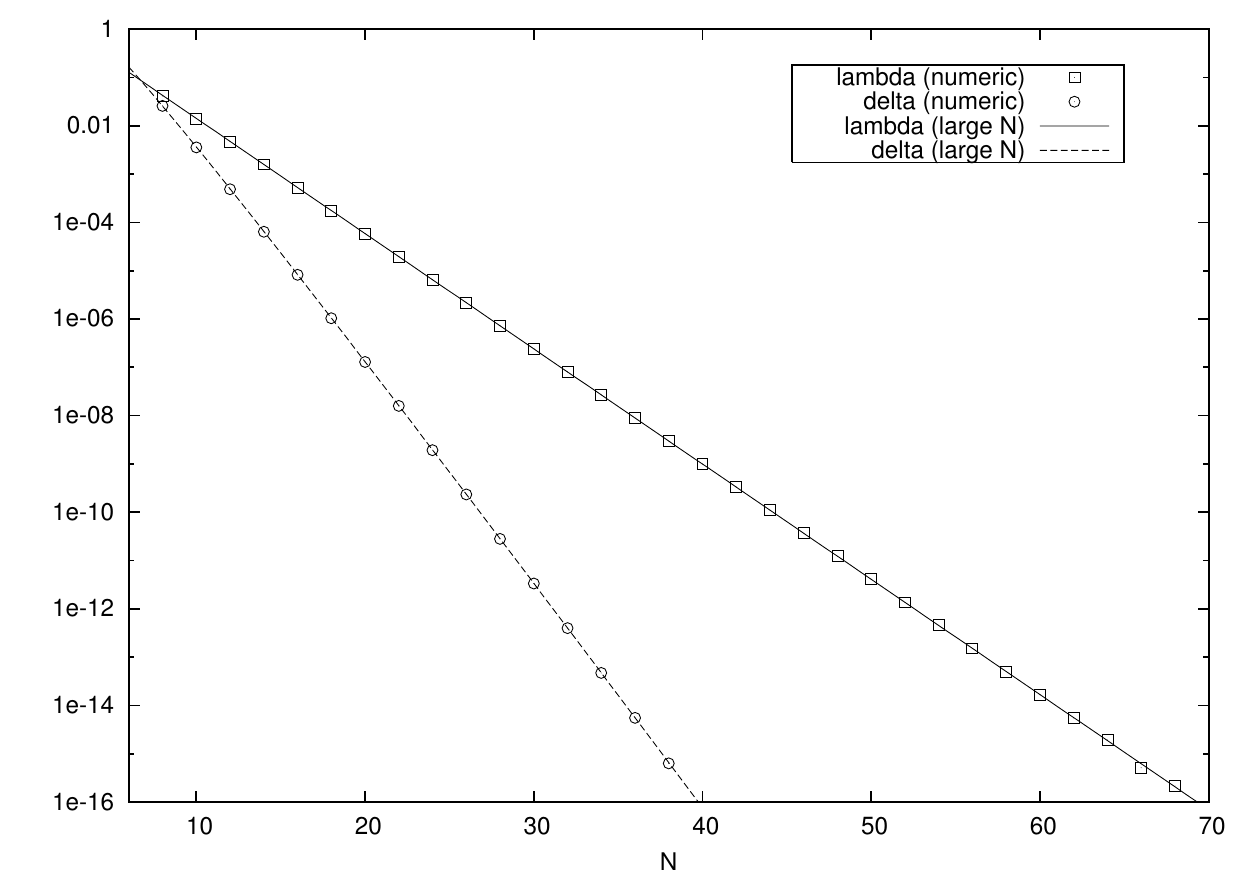}
\caption{Comparison of the values $\lambda$ (squares), $\delta$ (circles), in the absence of other roots, calculated by iteration of equation \eqref{eq:odd_sym_iter} and by the large-$N$ approximation \eqref{eq:odd_sym_large} (lines). The plot is limited to those values for which the iteration procedure remains within machine precision. It is seen that even for short chains, the large-$N$ approximation yields very good results.}
\label{fig:odd_symmetric}
\end{center}
\end{figure}

The full solution has no coinciding roots and the wave functions are regular Bethe wave functions.
The roots do, however, tend to grow very close as the chain length increases, leading to numerical problems: at more than $40$ sites machine precision is too low to find an acceptable result (see figure \ref{fig:odd_symmetric}). However, for smaller chains it is already clear that the values are exponentially decreasing.
For large $N$, inserting the assumption $\lambda \ll 1$, $\delta \ll 1$ in the Bethe equation, we note that by symmetry of the set $\set{\lambda_\beta}$, the contribution of the other roots is the real number $0 < F <1$ given by
\begin{align}
 F = \prod_{\beta}^{\lambda_\beta\not\in\set{\pm\lambda, \pm i(1+\delta)}}
\frac{\abs{\lambda_\beta}}{\sqrt{\lambda_\beta^2+1/4}} 
~.
\end{align}
This way we get
\begin{align}
\label{eq:odd_sym_large}
	\lambda &= \sqrt{\frac{12}{F}}\cdot 3^{-N/2}
&
	\delta &= \frac{24(N-1)}{F}\cdot 3^{-N}e^{-i\Phi}
~,
\end{align}
proving that the opposite real roots are exponentially close to each other, but can be pushed further 
apart in the presence of a macroscopic number of down spins (i.e. at low magnetic fields).

In this approximation, the reduced Bethe equations that must be satisfied by the remaining roots are
\begin{align}
N \arctan 2\lambda_j - 2 \arctan \lambda_j - \arctan \lambda_j/2 = \pi (J_j + \thalf) +
\hspace{0mm}\sum_{k=1}^{M-4}\hspace{0mm}
 \arctan(\lambda_j-\lambda_k) \modulo{\pi}
~,
\end{align}
if we order the rapidities such that the last four are the set $\set{\pm 0, \pm i}$.

The energy associated to the four roots $\set{\pm 0, \pm i}$ equals, for large $N$,
\begin{align}
	E_{\set{\pm 0, \pm i}} = -8/3
~.
\end{align}


\subsection{Singular pair states at $M=2$}
Here, besides superimposing pairs of even-length strings at the origin (which we will not discuss here;  their treatment
would extend the present section, and could be addressed by adopting a similar logic to that used for superimposed odd-length strings in the
previous section), we encounter a fundamentally different type of singular states, 
due to the presence of roots at the points $\pm i/2$. At this point, the Bethe equations are singular; moreover, the singularity in the kinetic phase is stronger than in the usual case for a string, $\bigoh{e^{\mp N \log\delta}}$ instead of $\bigoh{e^{\pm N}}$. This would suggest that this divergence cannot be countered in the usual way by the divergence in the scattering phase. 
Yet, if we carefully consider the way in which we take the limit $\lambda_\pm \goesto \pm i/2$, we shall see that we get a bona fide solution of the Bethe equations.

Note that the problem of singular states also arises in the XXZ model at `root of unity' values for the anisotropy parameter, as was shown by Fabricius and McCoy \cite{FabriciusMcCoy2001, FabriciusMcCoy2001b}: indeed, the exact complete $N$-strings of those articles correspond to the singular strings discussed here.

It was noted in [\onlinecite{Bethe1931,Essler1992}] that, in the $M=2$ sector, the wave function corresponding to the roots $\pm i/2$ is
\begin{align}
\label{eq:wave_fn_singular}
	\chi_{\pm i/2} (j_1, j_2) = (-1)^{j_1} \kron{j_1+1}{j_2} + (-1)^{j_2} \kron{j_1+N}{j_2+1}
~.
\end{align}

Here we have made the periodicity of the chain explicit; note that it is essential that $N$ is even for this state to exist. 
We shall see that if we take the limit $\lambda_\pm \goesto \pm i/2$ along the path prescribed by the Bethe equations, we recover the wave function \eqref{eq:wave_fn_singular}.

Consider $\lambda_\pm \definedby \epsilon \pm i(1+2\delta)/2$. We will take the limit $\epsilon\goesto 0$, $\delta\goesto 0$. As of yet, the signs of $\delta$ and $\epsilon$ are unspecified.
To first order, the kinetic phase satisfies
\begin{align}
	e^{ik_+} = e^{i\real{k_+}}e^{-\imag{k_+}} = \delta - i\epsilon
~,
\end{align}
 so that $e^{-\imag{k_+}}=\sqrt{\delta^2+\epsilon^2}$, $e^{i\real{k_+}}=  \sqrt{\frac{\delta-i\epsilon}{\delta+i\epsilon}}$. 
The scattering phase has (here, we denote $\Phi (k_+, k_-) = \phi (\lambda_+ - \lambda_-)$)
\begin{align}
	\tan \Phi(k_+, k_-)/2 = i(1-2e^{i\Phi(k_+, k_-)}) = \lambda_+ - \lambda_- = i(1+2\delta)
~;
\end{align} 
thus $e^{i\Phi} = -\delta$. Looking at the Bethe equations we now see
\begin{align}
	e^{ik_+N} &= -e^{i\Phi(k_+, k_-)}
\therefore
	(\delta - i\epsilon)^{N} = \delta
~.
\end{align}
so that we must have $\abs{\epsilon} \gg \abs{\delta}$; from which we deduce
\begin{align}
(-i)^N\epsilon^N &= \delta 
& 
e^{-\imag{k_+}} &= \abs{\epsilon}
&
e^{i\real{k_+}}&=  i
\end{align}
We see that, for $\epsilon$ and $\delta $  both to be real, we need $N$ even. Furthermore, 
\begin{align}
\label{eq:singular_narrow}
\sign\delta = (-1)^{N/2}.
 \forget{	
 \delta>0 &&\text{if $N\mod 4=0$}
\\
	\delta<0 &&\text{if $N\mod 4=2$}
}
\end{align}

Let us turn our attention to the wave function. We write 
$\chi_{(\pm i/2)} (j_1,j_2)  = \chi_{\sign\delta}^+ - \chi_{\sign\delta}^-$ where, for later convenience, we have defined
\begin{align}
\label{eq:def_psi_plus}
	\chi_a^+ (j_1, j_2) &\definedby 
	\lim_{\delta\goesto 0}^{\abs{\epsilon} = (a \delta)^{1/N}} e^{ik_+j_1+ik_-j_2}e^{i\Phi/2}
\\
\label{eq:def_psi_minus}
	\chi_a^- (j_1, j_2) &\definedby 
	\lim_{\delta\goesto 0}^{\abs{\epsilon}= (a \delta)^{1/N}} e^{ik_-j_1+ik_+j_2}e^{-i\Phi/2}
~.
\end{align}

Inserting the first-order values just found,
\begin{align}
	&\chi_a^\pm (j_1, j_2) = e^{i\real{k}(j_1+j_2)}e^{\mp \imag{k}(j_1-j_2)}e^{\mp i \Phi/2} = i^{j_1+j_2} (a\delta)^{\pm(j_1-j_2)/N}(-\delta)^{\pm 1/2} 
\notag
\end{align}
so that
\begin{align}
\label{eq:value_psi_plus}
	\chi_a^+ (j_1, j_2) &= - (a\delta)^{1/N}\delta^{-1/2}
	\left[ \inv a (-1)^{N/2 + j_2} \kron{j_1+N}{j_2+1} + \bigoh{\delta} \right]
\\
\label{eq:value_psi_minus}
	\chi_a^- (j_1, j_2) &= - (a\delta)^{1/N}\delta^{-1/2}
	\left[ (-1)^{j_1} \kron{j_1+1}{j_2} + \bigoh{\delta} \right]
\end{align}
We see that the prefactor, though divergent, is independent of position and therefore can be included 
in the normalization of the wave function $\chi_{\sign\delta}^+ - \chi_{\sign\delta}^-$. 
Due to \eqref{eq:singular_narrow}, it has the correct periodicity; we recover \eqref{eq:wave_fn_singular}. 

We did not have so specify the sign of $\epsilon$ in this derivation; it turns out that we can choose whether to approach the limit from the left or from the right half-plane.

We find the values of the quantum numbers when we consider the sum of the Bethe equations, viz.
\begin{align}
	\pi(J_++J_-) &= \lim_{\delta\goesto 0} N [\arctan(2\epsilon + i(1 + 2\delta)) + \arctan(2\epsilon - i(1 + 2\delta)) ]
\notag\\
		&= N \sign\epsilon \lim_{\delta\goesto 0}  \sign\delta \arctan\abs{\delta}^{-1+1/N} = N\pi/2 \modulo{N\pi}
	~,
\end{align}	
so that the Bethe quantum numbers for the singular state at $M=2$ are the half-integers
\begin{align}
J_+ &= \tquar[\pm N-(2-N \mod 4)] & J_- &= \tquar[\pm N+(2-N \mod 4)]
\end{align}
in agreement with \eqref{eq:singular_narrow}. The sign of these quantum numbers is not uniquely determined: it equals the sign of $\epsilon$ we chose in the limiting procedure.

Naturally, the Bethe--Takahashi quantum number corresponding to a single two-string at the origin is $I^{(2)} = 0$.


\subsection{Singular pair states at $M=3$}
For $M=3$, one singular state is already known: the $M=2$ state we just found, with an extra rapidity at infinity (i.e., momentum at zero).
However, another choice for the momentum that respects the lattice inversion symmetry is $k_3=\pi \inlinemodulo{2\pi}$ (i.e., $\lambda_3=0$). 
Let us use $\lambda_\pm = \epsilon \pm i(1+2\delta)/2$ and $\abs{\epsilon} \gg \abs{\delta}$ again.

The Bethe equation for $\lambda_+$ yields
\begin{align}
	\left[ 
		\frac{\epsilon + i\delta}{ \epsilon + i (1+\delta) }
	\right]^{N}
	 =
	  -e^{i\Phi_{13}}
	  \left[\frac{\delta}{1+\delta} \right]
\end{align}
As $\lambda_3\goesto 0, \lambda_+ \goesto i/2$, we have $e^{i\Phi_{13}} = e^{-i\Phi_{23}} = 1/3$, so that, to first order, $(-i\epsilon)^N=-\delta e^{i(\Phi_{13} - \Phi_{23})/2}$, and
 it turns out that we have to set $\sign\delta=(-1)^{1+N/2}$.

\forget{
Taking into account the fact that by symmetry $J_3 = 0$, the Bethe equation for $\lambda_3$ gives
\begin{align}
	N \arctan 2\lambda_0 = \arctan[ \lambda_0 - (\tfrac{1}{3}\abs\delta)^{1/N} - \tfrac{i}{2} -i\delta ] + \arctan[ \lambda_0 - (\tfrac{1}{3}\abs\delta)^{1/N} +\tfrac{i}{2} +i\delta ]
\notag
\end{align}
so that we find 
$\lambda_3 = \epsilon/(N-1) $.
}

Given this limiting procedure, we can now write the wave function as
\begin{align}
	\chi (j_1, j_2, j_3) \propto
		&\left[ \chi^+_a (j_1, j_2) - \chi^-_a (j_1, j_2) \right] e^{ik_3j_3} e^{i(\Phi_{13}+\Phi_{23})/2}
\\
		&+ \left[ \chi^+_a (j_2, j_3) - \chi^-_a (j_2, j_3) \right] e^{ik_3j_1} e^{-i(\Phi_{13}+\Phi_{23})/2}
\notag\\
		&+ \left[ \chi^+_a (j_1, j_3) e^{i(\Phi_{13}-\Phi_{23})/2} - \chi^-_a (j_1, j_3) e^{-i(\Phi_{13}-\Phi_{23})/2} \right] e^{ik_3j_2} 
\notag
~,
\end{align}
where $a = (-1)^{1+N/2}e^{i(\Phi_{13} - \Phi_{23})/2}$.

Because the $j$s are ordered, $\chi^+$ is zero unless its arguments are $j_1$ and $j_M$, and $\chi^-$ vanishes in that case only.
Using $e^{ik_3} = -1$ and the values from equations \eqref{eq:value_psi_plus}, \eqref{eq:value_psi_minus}, with the common prefactors divided out,
\begin{align}
	\chi^-_a (j_1,j_2) &= (-1)^{j_1} \kron{j_1+1}{j_2}
\\
	\chi^+_a (j_1,j_2) &= -e^{-i(\Phi_{13} - \Phi_{23})/2} (-1)^{j_2} \kron{j_1+N}{j_2+1} 
\notag
~,
\end{align}
we find that the wave function equals
\begin{align}
	\chi_{(\pm i/2, 0)} &(j_1, j_2, j_3) 
\\
	&\propto (-1)^{j_3} \chi^-_1 (j_1, j_2) + (-1)^{j_1} \chi^-_1 (j_2, j_3) + (-1)^{j_2} \chi^+_1 (j_1, j_3) 
\notag\\
	&\propto (-1)^{j_3+j_1} \kron{j_1+1}{j_2} + (-1)^{j_1+j_2}\kron{j_2+1}{j_3} + (-1)^{j_2+j_3} \kron{j_1+N}{j_3+1}
\notag
~.
\end{align}
Note that this wave function can be formed by simply creating a down spin of momentum $\pi$ on top of the $M=2$ singular state.

The Bethe quantum numbers are
\begin{align}
	J_+ &= \tquar[\pm N - (N \mod 4)] & J_0 &= 0 & J_- &= \tquar[\pm N+(N \mod 4)].
\end{align}
The Bethe--Takahashi quantum numbers are 
\begin{align}
	I^{(1)} &= 0 & I^{(2)} &= 0
~.
\end{align}


\subsection{Singular pair states at $M=4$}
Consider $M=4$ and $\lambda_\pm \definedby \epsilon \pm i(1+2\delta)$. Apart from the solutions we just found, extended with the appropriate number of infinite rapidities, we can find a few more. 

Note that for finite nonzero lambda, by symmetry, $\lambda_3 = -\lambda_4 \defines \lambda$. 
The Bethe equation for $\lambda_+$ now gives
\begin{align}
	\delta =  (-1)^{N/2} e^{-i(\Phi_{13} + \Phi_{14})} \epsilon^N
~.
\end{align}
By symmetry, $\Phi_{13} + \Phi_{14} = - \Phi_{23} - \Phi_{24} \defines\Phi/2$ and the wave function can be written
\begin{align}
	\chi&(j_1, j_2, j_3, j_4) =
\\
	&\quad  \chi^-_{ (-1)^{N/2}e^{i\Phi} }(j_1, j_2) \left[ e^{ik(j_3-j_4)} e^{i\Phi_{34}/2} -  e^{ik(j_4-j_3)} e^{-i\Phi_{34}/2} \right]
\notag\\
	&+ e^{i\Phi} \chi^+_
	{(-1)^{N/2}e^{i\Phi}}(j_1, j_4) \left[ e^{ik(j_2-j_3)} e^{i\Phi_{34}/2} -  e^{ik(j_3-j_2)} e^{-i\Phi_{34}/2} \right]
\notag\\
	&+ \chi^-_{(-1)^{N/2}e^{i\Phi}}(j_3, j_4) \left[e^{ik(j_1-j_2)} e^{i\Phi_{34}/2} -  e^{ik(j_2-j_1)} e^{-i\Phi_{34}/2} \right]
\notag\\
	&+ \chi^-_{(-1)^{N/2}e^{i\Phi}}(j_2, j_3) \left[ e^{ik(j_1-j_4)} e^{i\Phi_{34}/2}e^{i\Theta} -  e^{ik(j_4-j_1)} e^{-i\Phi_{34}/2}e^{-i\Theta}
	\right]
\notag 
\end{align}
where an extra phase factor has to be introduced for the terms in which the sites associated to $\lambda_3, \lambda_4$ surround the sites associated to the string,
\begin{align}
	\Theta \definedby  \thalf(\Phi_{13} + \Phi_{23}  - \Phi_{14} - \Phi_{24}) = 2 \arctan \tfrac{2}{3}\lambda + 2\arctan 2\lambda
~.
\end{align}
\forget{
Using the definitions of $\chi^\pm$, we see that $a$ and $e^{i\Theta}$ factors cancel, and
\begin{align}
	\chi(j_1, j_2, j_3, j_4) &\propto
		(-1)^{j_1} \kron{j_1+1}{j_2}\left[ e^{ik(j3-j4) - i\phi/2} - e^{ik(j4-j3) + i\phi/2} \right]
	\notag\\
		&+ (-1)^{j_3}\kron{j_3+1}{j_4}\left[ e^{ik(j1-j2) - i\phi/2} - e^{ik(j2-j1) + i\phi/2} \right]
	\notag\\
		&+ (-1)^{j_4} \kron{j_1+N}{j_4+1}\left[ e^{ik(j2-j3) - i\phi/2} - e^{ik(j3-j2) + i\phi/2} \right]
	\\
		&+ (-1)^{j_2} \kron{j_2+1}{j_3}\left[ e^{ik(j1-j4) - i(\phi+\Phi)/2} - e^{ik(j4-j1) + i(\phi+\Phi)/2} \right]
	\notag
~,
\end{align}
}

The Bethe equation for $\lambda$ can be reduced, using symmetry and the values $\lambda_\pm = \pm i/2$, to
\begin{align}
\label{eq:singular_M4_reduced_bethe}
 	(N-2) \arctan 2\lambda - \arctan \tfrac{2}{3} \lambda = \pi J_3
~.
\end{align}
For $J_3<(N-3)/2$, the solutions of \eqref{eq:singular_M4_reduced_bethe} are real and the Bethe--Takahashi quantum numbers are $I^{(2)}=0$ and $I^{(1)}_2 = -I^{(1)}_1 = J_3$.
For  $J_3 =(N-2)/2$, $\lambda$ is imaginary; this configuration can be identified as a {\em deviated four-string} with $I^{(4)}=0$.

The Bethe quantum numbers associated with $\lambda_\pm$ are the same as in the $M=2$ cases.


\subsection{Singular pair states at general $M$}
We will now generalise the approach of the last sections. Again, we only consider highest-weight states.
Consider a configuration of $M$ roots, two of which form a singular two-string, $\lambda\pm \definedby \epsilon \pm \frac{i}{2}(1-2\delta)$. The other roots must be distributed symmetrically; the total momentum is $\pi$ for even $M$, and $0$ for odd $M$. For ease of notation, let the number of non-singular nonzero roots be $\tilde M \definedby M - 2 -(M\mod 2)$; we'll indicate the particle at $k_0=\pi$ with the index $0$, and the two singular roots with $\pm$. The set is ordered $+,-,0,1 \ldots \tilde M$, 
such that we always have $k_a = -k_{\tilde M -a+1}$ for $a>0$.

The Bethe equation for $\lambda_+$ gives
\begin{align}
\label{eq:general_singular_limit}
	\delta 
	&= \epsilon^N  (-1)^{N/2} (-1)^M \prod_{\beta=0}^{\tilde M} e^{-i\Phi_{1\beta}}
~,
\end{align}
where $\Phi_{10}$ is understood to be zero if $M$ even.

Consider the Bethe wave function \eqref{eq:Bethe_Ansatz}. There will be one nonzero term with $\chi^+$: the one involving $\chi^+(j_1, j_M)$. This term has a nontrivial scattering with the singular string, equal to $e^{-i(\sum_{\beta=0}^M \Phi_{+\beta} -\Phi_{-\beta})}$; again, this cancels against the factor that arises taking the limit as above.

Another nontrivial scattering phase occurs when the sites $j_p, j_{p+1} = j_p+1$ are surrounded by a pair of sites associated with opposite momenta, yielding a phase
\begin{align}
	\Theta_\beta &\definedby \Phi(k_+,k_\beta)+\Phi(k_-,k_\beta) - \Phi(k_+,-k_\beta) -\Phi(k_-, -k_\beta)
\notag
\\
 	&= -2[\Phi(k_+, \abs{k_\beta})+\Phi(k_-, \abs{k_\beta})] 
~,
\end{align}
for $b\leq \tilde M/2$.

Moreover, only such permutations need be retained in the sum as make it possible for the sites involved in the singular complex to be adjacent.
With these considerations, the Bethe wave function becomes
\begin{align}
	\chi_{\pm i/2, \set{k}} (\set{j}) 
	&\propto 
	\sum_{\perm} (-1)^{\permsign{\perm}} \bigl[ 
		(-1)^{j_{\perm(+)}} \kron{\perm(-)}{1+\perm(+)} \kron{j_{\perm(+)}+1}{j_{1+\perm(+) }} 
\\
&\hspace{1cm}
	+ (-1)^{j_{1+ \perm(+)}} \kron{1+\perm(-)}{M+\perm(+)} \kron{j_{\perm(+)}+N}{j_{1+\perm(+)}+1} 
	\bigr]
	\times
\notag\\
&\hspace{-1cm}\times	
	e^{i\sum_{n=0}^{\tilde M} \left[k_n j_{\perm n} + \frac{1}{2} \sum^{\perm m > \perm n}_{0\leq m \leq\tilde M} \Phi_{mn}\right]}
\notag\\
&\hspace{-1cm}\times	
	e^{i\sum_{n=1}^{\tilde M/2} [\Phi(k_+, \abs{k_n}) + \Phi(k_-, \abs{k_n})][1+ \sign(\perm(+) -\perm n)\sign(\perm (\tilde M-n+1) - \perm(-))]}
\notag
	~.
\end{align}
Note that, for convenience, the permutations $\perm$ are the inverse of those in the earlier expression for the Bethe 
wave function \eqref{eq:Bethe_Ansatz}. The permutation $\perm$ is understood to be a map from $\set{+,-,0,1 \ldots \tilde M}$ to $\set{1 \ldots M}$.

We can make this expression slightly less ugly by splitting up the permutation $\perm$ such that $\perm = \perm_{+, a}\perm_{-, (a+1)} \permq_a$ where the permutation $\permq_a$ maps $\set{0,1\ldots \tilde M}$ to $\set{1\ldots a-1, a+2 \ldots M}$. Note that $\permsign{\permq_a} = \permsign{\perm}$; this separation is possible because the Kronecker symbols in the sum select only those permutations that map $+,-$ onto the neighbours $a, a+1$. 
Thus, the general Bethe wave function in the presence of a singular string reads
\begin{align}
\label{eq:general_singular}
	\chi_{\pm i/2, \set{k}} (\set{j}) 
	&\propto 
	\sum_a^{1\ldots N} \bigl[ 
		(-1)^{j_a} \kron{j_a+1}{j_{a+1}} 
		+ (-1)^{j_{a+1}} \kron{j_a+N}{j_{a+1}+1} 
	\bigr]\times
\notag\\
&\hspace{-1cm}
	\times\sum_{\permq_a} (-1)^{\permsign{\permq_a}}
	e^{i\sum_{n=1}^{\tilde M} \left[k_n j_{\permq n} + \frac{1}{2} \sum^{\permq m > \permq n}_{0\leq m\leq\tilde M} \Phi_{mn}\right]}
	\times
\\
	&\hspace{-1cm}
	\times
	e^{\frac{i}{2}\sum_{n=1}^{\tilde M} [\Phi(k_+, \abs{k_n}) + \Phi(k_-, \abs{k_n})][1 - \sign(a-\permq n)\sign(a - \permq(\tilde M-n+1))]}
\notag
	~.
\end{align}
The reduced Bethe equations that must be satisfied by the remaining roots are
\begin{align}
(N-1) \arctan 2\lambda_j - \arctan \tfrac{2}{3}\lambda_j = \pi J_j +
\hspace{0mm}\sum_{k=1}^{\tilde M}\hspace{0mm}
 \arctan(\lambda_j-\lambda_k) \modulo{\pi}
~.
\end{align}

The Bethe quantum numbers $J_+, J_-$ are those of the $M=2$ case (for $M$ even) or the $M=3$ case (for $M$ odd).


\subsection{Energy of singular pair}
In computing the energy contribution of the singular pair $\pm i/2$ we have to be careful to take the correct limit
\eqref{eq:general_singular_limit} arising from the Bethe equations. In this limit, the imaginary deviation is negligible compared to the real deviation, so that we may set $\lambda_\pm = \epsilon \pm i/2$. The energy contribution of the singular pair is then always 
\begin{align}
E_{\set{i/2, -i/2}} = -1
~.
\end{align}


\subsection{Validity of singular pair states}
It has been argued by Siddharthan \cite{Siddharthan1998} (on the basis of numerically calculated energies) and Noh et al. \cite{Noh2000} (by symmetry considerations) that singular pair states such as we just discussed, though they are solutions to the Bethe equations, do not (or not always) represent eigenstates of the Heisenberg model. However, we have shown above that they are valid solutions; moreover, we have checked the solutions of states found in this way against complete diagonalisation at $N=6$, $N = 8$ and $N=10$, and find perfect agreement.

As for the former article, the author finds six singular states for $N=6$, $M=3$, where only two are allowed by symmetry; it is therefore not surprising that four of those states are not eigenstates. The author fails to take the Bethe equations for the singular roots themselves correctly into account, and therefore finds too many solutions.
Considering the latter article, the situation with parity and translation symmetry is somewhat more subtle. It is considered in the next section, where it is used to prove that matrix elements of local operators with respect to the ground state and a singular pair state vanish.


\subsection{Form factors for singular pair states vanish}
\label{sub:vanish}
In the case of singular pair states, the reduced determinant expressions for correlators
\cite{Caux2005b} become degenerate. However, we can show that the form factors must vanish in this case with the following simple argument, based on their symmetry properties under lattice shifts and inversion.

\subsubsection{Translation symmetry}
All singular states have a symmetric rapidity distribution, which implies their total momentum must be either $0$ or $\pi$. The former implies symmetry under lattice shifts, and corresponds to the case with an odd number of finite roots aside from $\pm i/2$. The latter implies antisymmetry when shifting the lattice by one site, and corresponds to the case where there is an even number of such roots. 

\subsubsection{Parity symmetry}
Let us now turn our attention to the symmetry properties under lattice inversion (the parity operation). 
Under parity, the wave function $\chi_{\set\lambda} (j_1 \ldots j_M)$ is taken to $\chi_{\set\lambda} (N-j_M+1 \ldots N-j_1+1)$. Inserting this in the Bethe wave function, we get
\begin{align}
	\chi_{\set{k}} &(N-j_M+1 \ldots N-j_1+1 ) 
\notag\\
	&= 
 	e^{i \sum_\alpha k_{\alpha} (N+1)}
	A_0 \,\sum_{\perm} (-1)^{\permsign{\perm}}  e^{\half i \sum_{\alpha<\beta} \Phi(k_{\perm\alpha},\, k_{\perm\beta})} \,
	e^{i \sum_\alpha -k_{\perm\alpha} j_{M-\alpha}}
\\
	&= 
 	e^{i \sum_\alpha k_{\alpha} (N+1)}
	A_0 \,\sum_{\perm} (-1)^{\permsign{\perm}}  e^{\half i \sum_{\alpha<\beta} \Phi(-k_{\perm(M-\alpha)},\, -k_{\perm(M-\beta)})} \,
	e^{i \sum_\alpha -k_{\perm(M-\alpha)} j_{\alpha}}
	~,
\notag
\end{align}
Now let us set $k'_{M-\alpha} \definedby -k_\alpha$, so that 
\begin{align}
	\chi_{\set{k}} &(N-j_M+1 \ldots N-j_1+1 ) 
\\
	&= 
 	e^{i \sum_\alpha k_{\alpha} (N+1)}
	A_0 \,\sum_{\perm} (-1)^{\permsign{\perm}}  e^{\half i \sum_{\alpha<\beta} \Phi(k'_{\mathcal{V}\perm\mathcal{V}},\, k'_{\mathcal{V}\perm\mathcal{V}})} \,
	e^{i \sum_\alpha k'_{\mathcal{V}\perm\mathcal{V}} j_{\alpha}}
	~,
\notag
\end{align}
where $\mathcal{V}$ is the inversion permutation, such that $\mathcal{V}\perm\mathcal{V} = M-\perm(M-\alpha)$.
Realising that $\permsign{\mathcal{V}\perm\mathcal{V}} = \permsign{\perm}$, we see that we recover the Bethe wave function for the set of momenta $\set{k'}$, with a prefactor $e^{i \sum_\alpha k_{\alpha} (N+1)} = e^{i \sum_\alpha k_{\alpha}}$. Therefore, a Bethe wave function with $\set{k} = \set{-k}$ is either symmetric (if $\sum_\alpha k_{\alpha} = 0$) or antisymmetric (if $\sum_\alpha k_{\alpha} = \pi$) under parity; conversely, if a state $\chi_{\set{k}}$ is an eigenfunction of parity at eigenvalue $v$, we must have $\chi_{\set{k}}= v\chi_{\set{-k}}$.

Consider, however, a singular Bethe wave function; take for instance $\chi_{\pm i/2} (j_1,j_2) = (-1)^{j_1} \kron{j_1}{j_2} + (-1)^{j_2} \kron{j1+N}{j2+1}$. It is easily seen that this state is symmetric under parity, but antisymmetric under a single-site shift.
 In [\onlinecite{Noh2000}] this is taken to be a contradiction, implying that singular states cannot be Bethe states. However, even though $\set{\lambda} = \set{-\lambda}$, the momenta for the singular pair are $k_{\pm} = \pi/2 \pm i\infty$; the limit that needs to be taken to arrive as this point is such that at no point $\lambda_- = -\lambda_+$ except at the limit itself. The opposite momenta $k_{\pm} = -\pi/2 \pm i\infty$ yield, in the limit, the same wave function; but $\set{k} \neq \set{-k}$ and therefore the assumption leading to the symmetry relation above is not satisfied for singular Bethe states.

Applying the above argument to the singular-state wave function of \eqref{eq:general_singular}, we see that lattice inversion takes the {\em finite} momenta to their opposites; but the part of the wave function corresponding to the singular pair is taken to itself. (Note that the extra scattering factor for opposite momenta surrounding a pair does not change, as the property of surrounding something is invariant under parity).
Thus the eigenvalue under parity from a singular state is $-e^{i \sum_\alpha k_{\alpha}}$. Indeed, for a general singular state, symmetry under shifts implies antisymmetry under parity and vice versa. 

\subsubsection{Symmetry and form factors}
For the general study of dynamics using all important excited states \cite{Caux2005, Caux2005b, CauxHagemans2006} 
it is essential to be able to determine all required matrix elements of local spin operators (form factors). 
It is easy to show that form factors of singular states with the ground state vanish. Consider an even singular state, i.e. of momentum $\pi$. The ground state for even $M$ has momentum $0$. Therefore, the only possibly nonvanishing form factor operates at momentum $\pi$,
\begin{align}
	F^\alpha_\pi (\text{GS},\set{\pm i/2, \lambda}) = \bra{\text{GS}}{\sum_j (-1)^j S^\alpha_j}\ket{\set{\pm i/2, \lambda}}.
\end{align}
As we have shown, the state $\ket{\set{\pm i/2, \lambda}}$ is symmetric under parity. Since $N$ is even, the state $\sum_j (-1)^j S^\alpha_j \ket{\set{\pm i/2, \lambda}}$ is antisymmetric under parity. But the ground state $\ket{\text{GS}}$ is symmetric; therefore, their overlap (which is the form factor) must be zero. 
The converse argument holds when we consider odd singular states; again, the form factor is zero.




\section{Conclusion}
In this paper we have discussed a method allowing us to find an extended class of solutions of the Bethe equations in the complex plane, 
by calculating deviations to the string-hypothesis solution. For fixed $M$ at increasing $N$, the fraction of solutions that 
cannot be found decreases algebraically with the number of sites and factorially with the number of magnons. The behaviour of 
the number of missing solutions suggests that collapsing pairs are the only source of failure of the string picture in this regime. 
The average deviation of the string hypothesis is found to decrease with $N$, but it is not exponential in all cases.  
If the number of magnons becomes macroscopic, {\it i.e.} when the field becomes small, we expect that the method becomes more
difficult to implement, as can be expected from general arguments (see {\it e.g.} [\onlinecite{TsvelikWiegmann1983}]).

We have also shown that singular states exist whenever a symmetric configuration includes an even string at the origin, leading to a singularity in the Bethe equations. These states are generalisations of the $M=2$ singular state that had been known for a long time. We also show that, in contrast to what is claimed in the literature, these states can be seen as legitimate solutions to the Bethe equations, as long as the limit is taken in the correct way. We have also shown that all states in this class have zero form factors for local spin operators, and can therefore be ignored in the calculation of zero-temperature correlation functions of local spins.  We have not yet investigated cases in which either a four- or higher even string is present at the origin in
a symmetric state, nor the case of superimposed pairs of even or odd strings at the origin.  
Yet another class of special solutions arises when a symmetric configuration includes two odd-length strings at the origin. Both of these last two
situations lead to solutions with pairs of roots that are exponentially close to each other as $N$ becomes large, without violating the 
exclusion principle. Form factors and norms for these states exist and are nonzero, but are hard to calculate due to the exponential degeneracy. 
We leave all of these for future work.

In summary, we have presented sets of equations allowing to obtain eigenstates of the Heisenberg chain beyond
the traditional string hypothesis.  These results are of importance in particular for the calculation of dynamical correlation
functions of finite chains, where form factors depend sensitively on deviations from pure strings.  On a more formal
level, although the state counting can be categorized using partitioning of rapidities in pure strings, the actual solutions
to the Bethe equations can differ substantially from the string hypothesis, and can do so in very elaborate ways 
(especially for the higher strings which we consider here).  We hope that the present work will provide stimulation
for an eventual more faithful and representative classification of solutions to the Bethe equations.








\section{Acknowledgements}
R. H. and J.-S. C. acknowledge support from the Stichting voor Fundamenteel Onderzoek der Materie (FOM)
of the Netherlands.



\appendix

\section{Complete Bethe Ansatz solutions of finite chains}
\label{appendix_finite}

In tables \ref{tab:N8M1}--\ref{tab:N8M4} we give the rapidities for all highest-weight eigenstates of the isotropic Heisenberg chain at $8$ sites, except the reference state (which has no rapidities). Lower-weight states can be formed by adding the appropriate number of infinite rapidities. 
Similarly, tables \ref{tab:N10M1}--\ref{tab:N10M5b} contain the results for $10$ sites.

At this size, the model is easily completely diagonalisable and therefore we do not need the Bethe Ansatz to solve it. 
However, while constructing the complete solution from the Bethe Ansatz solutions is easy for this number of particles, mapping the exactly diagonalised wave functions to solutions of the Bethe equations is not. Therefore, we believe it valuable to have an exhaustive list at hand.
The solutions are found by the methods of section \ref{sec:deviations}. The quantum numbers are determined from the rapidities, thereby providing a valuable check on their accuracy as solutions to the Bethe equations. We have also checked these solutions against a complete diagonalisation, and found full agreement.  One thing one can easily see is that these solutions neatly satisfy the classification in terms of the string hypothesis: the states can be associated one-to-one  with string states, with small but significant deviations, and with quantum numbers that satisfy the bound \eqref{eq:quantum_number_bound}.
\begin{table}
{\tiny
\begin{center}
\begin{tabular}{r|r|r|r|r}
&\multicolumn{1}{r|}{$2J$}&\multicolumn{1}{r|}{$2I_j$}&\multicolumn{1}{r|}{$\lambda$}&$E_0$\\
\hline
\hline
1. &	$0$ &	$0_1$ &	$0$ &	$-2$ \\
2. &	$-2$ &	$-2_1$ &	$-0.207106781186548$ &	$-1.70710678118655$ \\
3. &	$2$ &	$2_1$ &	$0.207106781186548$ &	$-1.70710678118655$ \\
4. &	$-4$ &	$-4_1$ &	$-0.5$ &	$-1$ \\
5. &	$4$ &	$4_1$ &	$0.5$ &	$-1$ \\
6. &	$-6$ &	$-6_1$ &	$-1.20710678118655$ &	$-0.292893218813452$ \\
7. &	$6$ &	$6_1$ &	$1.20710678118655$ &	$-0.292893218813452$ \\
\hline
\end{tabular}
\end{center}
}
\caption{Highest-weight Bethe Ansatz solutions for $N=8$, $M=1$}
\label{tab:N8M1}
\end{table}

\begin{table}
{\tiny
\begin{center}
\begin{tabular}{r|rr|rr|rr|r} 
&\multicolumn{2}{r|}{$2J$}&\multicolumn{2}{r|}{$2I_j$}&\multicolumn{2}{r|}{$\lambda$}&$E_0$\\
\hline
\hline
1. &	$-1$ &	$1$ &	$-1_1$ &	$1_1$ &	$-0.114121737195075$ &	$0.114121737195075$ &	$-3.80193773580484$ \\
2. &	$-1$ &	$-3$ &	$-1_1$ &	$-3_1$ &	$-0.0820036908620831$ &	$-0.359101081059097$ &	$-3.26703509836137$ \\
3. &	$1$ &	$-3$ &	$1_1$ &	$-3_1$ &	$0.130438947345799$ &	$-0.378439593963216$ &	$-3.14412280563537$ \\
4. &	$-1$ &	$3$ &	$-1_1$ &	$3_1$ &	$-0.130438947345799$ &	$0.378439593963216$ &	$-3.14412280563537$ \\
5. &	$1$ &	$3$ &	$1_1$ &	$3_1$ &	$0.082003690862083$ &	$0.359101081059097$ &	$-3.26703509836137$ \\
6. &	$-1$ &	$-5$ &	$-1_1$ &	$-5_1$ &	$-0.0539607683954299$ &	$-0.914799098278114$ &	$-2.43701602444882$ \\
7. &	$1$ &	$-5$ &	$1_1$ &	$-5_1$ &	$0.155070403217499$ &	$-0.949571172389941$ &	$-2.25865202250415$ \\
8. &	$-1$ &	$5$ &	$-1_1$ &	$5_1$ &	$-0.155070403217499$ &	$0.949571172389941$ &	$-2.25865202250415$ \\
9. &	$1$ &	$5$ &	$1_1$ &	$5_1$ &	$0.0539607683954298$ &	$0.914799098278114$ &	$-2.43701602444882$ \\
10. &	$-3$ &	$3$ &	$-3_1$ &	$3_1$ &	$-0.398736694441202$ &	$0.398736694441202$ &	$-2.44504186791263$ \\
11. &	$-3$ &	$-5$ &	$-3_1$ &	$-5_1$ &	$-0.288675134594813$ &	$-0.866025403784438$ &	$-2$ \\
12. &	$-3$ &	$5$ &	$-3_1$ &	$5_1$ &	$-0.428405998277582$ &	$0.985143205784529$ &	$-1.56298397555118$ \\
13. &	$3$ &	$-5$ &	$3_1$ &	$-5_1$ &	$0.428405998277582$ &	$-0.985143205784529$ &	$-1.56298397555118$ \\
14. &	$3$ &	$5$ &	$3_1$ &	$5_1$ &	$0.288675134594813$ &	$0.866025403784438$ &	$-2$ \\
15. &	$-5$ &	$5$ &	$-5_1$ &	$5_1$ &	$-1.03826069828617$ &	$1.03826069828617$ &	$-0.753020396282533$ \\
\hline
16. &	$3$ &	$5$ &	\multicolumn{2}{c|}{$0_2$} &	\multicolumn{2}{c|}{$\pm 0.5i$} &	$-1$ \\
17. &	$-5$ &	$-5$ &	\multicolumn{2}{c|}{$-2_2$} &	\multicolumn{2}{c|}{$-0.415344339607922\pm 0.499530117286392i$} &	$-0.855877194364632$ \\
18. &	$5$ &	$5$ &	\multicolumn{2}{c|}{$2_2$} &	\multicolumn{2}{c|}{$0.415344339607923\pm 0.499530117286392i$} &	$-0.855877194364631$ \\
19. &	$-7$ &	$-5$ &	\multicolumn{2}{c|}{$-4_2$} &	\multicolumn{2}{c|}{$-0.951136103880144\pm 0.544496295465675i$} &	$-0.474312879134482$ \\
20. &	$5$ &	$7$ &	\multicolumn{2}{c|}{$4_2$} &	\multicolumn{2}{c|}{$0.951136103880144\pm 0.544496295465675i$} &	$-0.474312879134482$ \\
\hline
\end{tabular}
\end{center}
}
\caption{Highest-weight Bethe Ansatz solutions for $N=8$, $M=2$. There are $15$ states with all real solutions, and $5$ with a single two-string. }
\label{tab:N8M2}
\end{table}


\begin{table}
{\tiny
\begin{center}
\begin{tabular}{r|rrr|rrr|rrr|r}
\hline
\hline
1. &	$0$ &	$-2$ &	$2$ &	$0_1$ &	$-2_1$ &	$2_1$ &	$0$ &	$-0.263913376890051$ &	$0.263913376890051$ &	$-5.12841906384458$ \\
2. &	$0$ &	$-2$ &	$-4$ &	$0_1$ &	$-2_1$ &	$-4_1$ &	$0.0537192218681322$ &	$-0.193583000298505$ &	$-0.650848824168377$ &	$-4.45873850889483$ \\
3. &	$0$ &	$2$ &	$-4$ &	$0_1$ &	$2_1$ &	$-4_1$ &	$0.0238176981350052$ &	$0.288834213916909$ &	$-0.720503071221302$ &	$-4.14514837392072$ \\
4. &	$-2$ &	$2$ &	$-4$ &	$-2_1$ &	$2_1$ &	$-4_1$ &	$-0.209705666535535$ &	$0.306323520147911$ &	$-0.681661096925392$ &	$-3.85463767971846$ \\
5. &	$0$ &	$-2$ &	$4$ &	$0_1$ &	$-2_1$ &	$4_1$ &	$-0.0238176981350052$ &	$-0.288834213916909$ &	$0.720503071221302$ &	$-4.14514837392072$ \\
6. &	$0$ &	$2$ &	$4$ &	$0_1$ &	$2_1$ &	$4_1$ &	$-0.0537192218681322$ &	$0.193583000298505$ &	$0.650848824168377$ &	$-4.45873850889483$ \\
7. &	$-2$ &	$2$ &	$4$ &	$-2_1$ &	$2_1$ &	$4_1$ &	$-0.306323520147911$ &	$0.209705666535536$ &	$0.681661096925392$ &	$-3.85463767971846$ \\
8. &	$0$ &	$-4$ &	$4$ &	$0_1$ &	$-4_1$ &	$4_1$ &	$0$ &	$-0.763017839489689$ &	$0.763017839489689$ &	$-3.20163967572341$ \\
9. &	$-2$ &	$-4$ &	$4$ &	$-2_1$ &	$-4_1$ &	$4_1$ &	$-0.232643790002621$ &	$-0.723239730965897$ &	$0.793819433881308$ &	$-2.85892354971026$ \\
10. &	$2$ &	$-4$ &	$4$ &	$2_1$ &	$-4_1$ &	$4_1$ &	$0.232643790002621$ &	$-0.793819433881308$ &	$0.723239730965896$ &	$-2.85892354971026$ \\
\hline
11. &	$0$ &	$4$ &	$4$ &	$0_1$ &	\multicolumn{2}{c|}{$0_2$} &	$0$ &	\multicolumn{2}{c|}{$\pm 0.5i$} &	$-3$ \\
12. &	$0$ &	$-6$ &	$-4$ &	$0_1$ &	\multicolumn{2}{c|}{$-2_2$} &	$0.0888359485197147$ &	\multicolumn{2}{c|}{$-0.620939972021564\pm 0.511033534711325i$} &	$-2.62837842688306$ \\
13. &	$0$ &	$4$ &	$6$ &	$0_1$ &	\multicolumn{2}{c|}{$2_2$} &	$-0.0888359485197147$ &	\multicolumn{2}{c|}{$0.620939972021564\pm 0.511033534711325i$} &	$-2.62837842688306$ \\
14. &	$-2$ &	$4$ &	$4$ &	$-2_1$ &	\multicolumn{2}{c|}{$0_2$} &	$-0.278658967238795$ &	\multicolumn{2}{c|}{$0.116265462834909\pm 0.499999922966273i$} &	$-2.51268320306387$ \\
15. &	$2$ &	$-4$ &	$-4$ &	$2_1$ &	\multicolumn{2}{c|}{$0_2$} &	$0.278658967238795$ &	\multicolumn{2}{c|}{$-0.116265462834909\pm 0.499999922966273i$} &	$-2.51268320306387$ \\
16. &	$-4$ &	$4$ &	$4$ &	$-4_1$ &	\multicolumn{2}{c|}{$0_2$} &	$-0.771265809851148$ &	\multicolumn{2}{c|}{$0.213409859292069\pm 0.499994150006002i$} &	$-1.54839403704422$ \\
17. &	$4$ &	$-4$ &	$-4$ &	$4_1$ &	\multicolumn{2}{c|}{$0_2$} &	$0.771265809851148$ &	\multicolumn{2}{c|}{$-0.213409859292069\pm 0.499994150006002i$} &	$-1.54839403704422$ \\
18. &	$-2$ &	$-6$ &	$-4$ &	$-2_1$ &	\multicolumn{2}{c|}{$-2_2$} &	$-0.139811316778799$ &	\multicolumn{2}{c|}{$-0.550392917633605\pm 0.50687006912391i$} &	$-2.59696828323732$ \\
19. &	$2$ &	$-6$ &	$-4$ &	$2_1$ &	\multicolumn{2}{c|}{$-2_2$} &	$0.347810384779931$ &	\multicolumn{2}{c|}{$-0.673905192389966\pm 0.514426127068346i$} &	$-2$ \\
20. &	$-4$ &	$-6$ &	$-4$ &	$-4_1$ &	\multicolumn{2}{c|}{$-2_2$} &	$-0.595672174122518$ &	\multicolumn{2}{c|}{$-0.314924940488343\pm 0.500176262487235i$} &	$-1.73454728842451$ \\
21. &	$4$ &	$-6$ &	$-4$ &	$4_1$ &	\multicolumn{2}{c|}{$-2_2$} &	$0.865745241274482$ &	\multicolumn{2}{c|}{$-0.740655728477336\pm 0.519219952109156i$} &	$-1.10730664234478$ \\
22. &	$-2$ &	$4$ &	$6$ &	$-2_1$ &	\multicolumn{2}{c|}{$2_2$} &	$-0.347810384779931$ &	\multicolumn{2}{c|}{$0.673905192389965\pm 0.514426127068346i$} &	$-2$ \\
23. &	$2$ &	$4$ &	$6$ &	$2_1$ &	\multicolumn{2}{c|}{$2_2$} &	$0.139811316778799$ &	\multicolumn{2}{c|}{$0.550392917633605\pm 0.50687006912391i$} &	$-2.59696828323731$ \\
24. &	$-4$ &	$4$ &	$6$ &	$-4_1$ &	\multicolumn{2}{c|}{$2_2$} &	$-0.865745241274482$ &	\multicolumn{2}{c|}{$0.740655728477336\pm 0.519219952109156i$} &	$-1.10730664234478$ \\
25. &	$4$ &	$4$ &	$6$ &	$4_1$ &	\multicolumn{2}{c|}{$2_2$} &	$0.595672174122519$ &	\multicolumn{2}{c|}{$0.314924940488342\pm 0.500176262487235i$} &	$-1.73454728842451$ \\
\hline
26. &	$6$ &	$0$ &	$10$ &	\multicolumn{3}{c|}{$0_3$} &	\multicolumn{2}{c}{$\pm 1.00092282114108i$} &	$0$ &	$-0.669941260432018$ \\
27. &	$-8$ &	$-4$ &	$-6$ &	\multicolumn{3}{c|}{$-2_3$} &	\multicolumn{2}{c}{$-0.615758454071295\pm 0.988145332430968i$} &	$-0.631196681886563$ &	$-0.554274006757972$ \\
28. &	$6$ &	$4$ &	$8$ &	\multicolumn{3}{c|}{$2_3$} &	\multicolumn{2}{c}{$0.615758454071295\pm 0.988145332430968i$} &	$0.631196681886563$ &	$-0.554274006757972$ \\
\hline
\end{tabular}
\end{center}
}
\caption{Highest-weight Bethe Ansatz solutions for $N=8$, $M=3$. There are $10$ solutions with all reals, $15$ with one two-string, and $3$ with a single three-string. We give both Bethe $J$ and Bethe--Takahashi $I$ quantum numbers. The latter are subscripted by the length of the string for which they are quantum numbers.}
\label{tab:N8M3}
\end{table}

\begin{table}
{\tiny
\begin{center}
\begin{tabular}{r|rrrr|rrrr|rrrr|r}
 &\multicolumn{4}{r|}{$2J$}&\multicolumn{4}{r|}{$2I_j$}&\multicolumn{4}{r|}{$\lambda$}&$E_0$\\
\hline
\hline
1. &	$-1$ &	$1$ &	$-3$ &	$3$ &	$-1_1$ &	$1_1$ &	$-3_1$ &	$3_1$ &	$-0.129472946374929$ &	$0.129472946374929$ &	$-0.525012102223667$ &	$0.525012102223667$ &	$-5.65109340893718$ \\
\hline
2. &	$-1$ &	$1$ &	$3$ &	$5$ &	$-1_1$ &	$1_1$ &	\multicolumn{2}{c|}{$0_2$} &	$-0.142469067830567$ &	$0.142469067830566$ &	\multicolumn{2}{c|}{$\pm 0.5i$} &	$-4.69962814827532$ \\
3. &	$-1$ &	$-3$ &	$3$ &	$5$ &	$-1_1$ &	$-3_1$ &	\multicolumn{2}{c|}{$0_2$} &	$-0.147012611196141$ &	$-0.557070238574442$ &	\multicolumn{2}{c|}{$0.352041424885292\pm 0.500558169643331i$} &	$-3.61803398874989$ \\
4. &	$1$ &	$-3$ &	$3$ &	$5$ &	$1_1$ &	$-3_1$ &	\multicolumn{2}{c|}{$0_2$} &	$0.121186177969172$ &	$-0.571611177186439$ &	\multicolumn{2}{c|}{$0.225212499608634\pm 0.500028862163533i$} &	$-3.70710678118655$ \\
5. &	$-1$ &	$3$ &	$-5$ &	$-3$ &	$-1_1$ &	$3_1$ &	\multicolumn{2}{c|}{$0_2$} &	$-0.121186177969172$ &	$0.571611177186439$ &	\multicolumn{2}{c|}{$-0.225212499608634\pm 0.500028862163533i$} &	$-3.70710678118655$ \\
6. &	$1$ &	$3$ &	$-5$ &	$-3$ &	$1_1$ &	$3_1$ &	\multicolumn{2}{c|}{$0_2$} &	$0.147012611196141$ &	$0.557070238574442$ &	\multicolumn{2}{c|}{$-0.352041424885292\pm 0.500558169643331i$} &	$-3.61803398874989$ \\
7. &	$-3$ &	$3$ &	$3$ &	$5$ &	$-3_1$ &	$3_1$ &	\multicolumn{2}{c|}{$0_2$} &	$-0.563825262393496$ &	$0.563825262393496$ &	\multicolumn{2}{c|}{$\pm 0.5i$} &	$-2.76087672174345$ \\
\hline
8. &	$1$ &	$7$ &	$-1$ &	$-7$ &	$0_1$ &	\multicolumn{3}{c|}{$0_3$} &	$-0.041309127524556$ &	\multicolumn{2}{c}{$\pm 1.02570508123074i$} &	$0.041309127524556$ &	$-2.72610944503578$ \\
9. &	$-1$ &	$7$ &	$-1$ &	$9$ &	$-2_1$ &	\multicolumn{3}{c|}{$0_3$} &	$-0.244333193771166$ &	\multicolumn{2}{c}{$0.0802730431898699\pm 1.00558827395993i$} &	$0.0837871073914291$ &	$-2.29289321881345$ \\
10. &	$1$ &	$-9$ &	$1$ &	$-7$ &	$2_1$ &	\multicolumn{3}{c|}{$0_3$} &	$0.244333193771166$ &	\multicolumn{2}{c}{$-0.0802730431898705\pm 1.00558827395993i$} &	$-0.0837871073914297$ &	$-2.29289321881345$ \\
11. &	$-3$ &	$5$ &	$3$ &	$7$ &	$-4_1$ &	\multicolumn{3}{c|}{$0_3$} &	$-0.669122922881511$ &	\multicolumn{2}{c}{$0.224281457794713\pm 1.00224727650661i$} &	$0.220560007292084$ &	$-1.38196601125011$ \\
12. &	$3$ &	$-7$ &	$-3$ &	$-5$ &	$4_1$ &	\multicolumn{3}{c|}{$0_3$} &	$0.669122922881511$ &	\multicolumn{2}{c}{$-0.224281457794714\pm 1.00224727650661i$} &	$-0.220560007292085$ &	$-1.38196601125011$ \\
\hline
13. &	$5$ &	$3$ &	$5$ &	$7$ &	\multicolumn{4}{c|}{$0_4$} &	\multicolumn{2}{c}{$\pm 1.55612650357705i$} &	\multicolumn{2}{c|}{$\pm 0.5i$} &	$-0.539495129981236$ \\
\hline
14. &	$-5$ &	$-3$ &	$3$ &	$5$ &	\multicolumn{2}{c}{$-1_2$} &	\multicolumn{2}{c|}{$1_2$} &	\multicolumn{2}{c}{$-0.463264727589031\pm 0.502293853569903i$} &	\multicolumn{2}{c|}{$0.463264727589031\pm 0.502293853569903i$} &	$-1.62279714602704$ \\
\hline
\end{tabular}
\end{center}
}
\caption{Highest-weight Bethe Ansatz solutions for $N=8$, $M=4$. There is $1$ state with all real solutions, $6$ with one two-string, $5$ with one three-string, $1$ single four-string, and $1$ with two two-strings.}
\label{tab:N8M4}
\end{table}



\begin{table}
{\tiny
\begin{center}
\begin{tabular}{r|r|r|r|r} 
&\multicolumn{1}{r|}{$2J$}&\multicolumn{1}{r|}{$2I_j$}&\multicolumn{1}{r|}{$\lambda$}&$E_0$\\
\hline
\hline
1. &    $0$ &   $0_1$ & $0$ &   $-2$ \\
2. &    $-2$ &  $-2_1$ &        $-0.162459848116453$ &  $-1.80901699437495$ \\
3. &    $2$ &   $2_1$ & $0.162459848116453$ &   $-1.80901699437495$ \\
4. &    $-4$ &  $-4_1$ &        $-0.36327126400268$ &   $-1.30901699437495$ \\
5. &    $4$ &   $4_1$ & $0.36327126400268$ &    $-1.30901699437495$ \\
6. &    $-6$ &  $-6_1$ &        $-0.688190960235587$ &  $-0.690983005625053$ \\
7. &    $6$ &   $6_1$ & $0.688190960235587$ &   $-0.690983005625053$ \\
8. &    $-8$ &  $-8_1$ &        $-1.53884176858763$ &   $-0.190983005625053$ \\
9. &    $8$ &   $8_1$ & $1.53884176858763$ &    $-0.190983005625053$ \\
\hline
\end{tabular}
\end{center}
\caption{Highest-weight Bethe Ansatz solutions for $N=10$, $M=1$.}
\label{tab:N10M1}
}
\end{table}

\begin{table}
{\tiny
\begin{center}
\begin{tabular}{r|rr|rr|rr|r} 
&\multicolumn{2}{r|}{$2J$}&\multicolumn{2}{r|}{$2I_j$}&\multicolumn{2}{r|}{$\lambda$}&$E_0$\\
\hline
\hline
1. &    $-1$ &  $1$ &   $-1_1$ &    $1_1$ & 		$-0.0881634903542325$ & $0.0881634903542325$ &  $-3.87938524157182$ \\
2. &    $-1$ &  $-3$ &  $-1_1$ &    $-3_1$ &        $-0.069203328842458$ &  $-0.267198865360809$ &  $-3.51812424513935$ \\
3. &    $1$ &   $-3$ &  $1_1$ & 	$-3_1$ &        $0.097718323279698$ &   $-0.277820084547718$ &  $-3.45461156846711$ \\
4. &    $-1$ &  $3$ &   $-1_1$ &    $3_1$ & 		$-0.097718323279698$ &  $0.277820084547718$ &   $-3.45461156846711$ \\
5. &    $1$ &   $3$ &   $1_1$ & 	$3_1$ & 		$0.0692033288424579$ &  $0.267198865360809$ &   $-3.51812424513935$ \\
6. &    $-1$ &  $-5$ &  $-1_1$ &    $-5_1$ &        $-0.0559079379624147$ & $-0.547952407065598$ &  $-2.88397803317447$ \\
7. &    $1$ &   $-5$ &  $1_1$ & 	$-5_1$ &        $0.109874636222197$ &   $-0.563039149833424$ &  $-2.78968334011566$ \\
\hline
\end{tabular}
\end{center}
}
\caption{Highest-weight Bethe Ansatz solutions for $N=10$, $M=2$. There are $28$ states with only real roots and $7$ two-strings.}
\label{tab:N10M2}
\end{table}

\begin{table}
\hspace{-17mm}
{\tiny
\begin{tabular}{r|rrr|rrr|rrr|r}
 &\multicolumn{3}{r|}{$2J$}&\multicolumn{3}{r|}{$2I_j$}&\multicolumn{3}{r|}{$\lambda$}&$E_0$\\
 \hline
 \hline
1. &    $0$ &   $-2$ &  $2$ &   $0_1$ & $-2_1$ &        $2_1$ & $0$ &   $-0.194145912742908$ &  $0.194145912742908$ &   $-5.47593186905444$ \\
2. &    $0$ &   $-2$ &  $-4$ &  $0_1$ & $-2_1$ &        $-4_1$ &        $0.0304484059690835$ &  $-0.158812988572359$ &  $-0.417525851428647$ &  $-4.98766285797636$ \\
3. &    $0$ &   $2$ &   $-4$ &  $0_1$ & $2_1$ & $-4_1$ &        $0.0119052857672312$ &  $0.206226507953237$ &   $-0.446703652043529$ &  $-4.82033462714144$ \\
4. &    $-2$ &  $2$ &   $-4$ &  $-2_1$ &        $2_1$ & $-4_1$ &        $-0.16860988033085$ &   $0.216339433204765$ &   $-0.431009948231322$ &  $-4.62780396148396$ \\
5. &    $0$ &   $-2$ &  $4$ &   $0_1$ & $-2_1$ &        $4_1$ & $-0.0119052857672312$ & $-0.206226507953237$ &  $0.446703652043529$ &   $-4.82033462714144$ \\
6. &    $0$ &   $2$ &   $4$ &   $0_1$ & $2_1$ & $4_1$ & $-0.0304484059690835$ & $0.158812988572359$ &   $0.417525851428647$ &   $-4.98766285797636$ \\
7. &    $-2$ &  $2$ &   $4$ &   $-2_1$ &        $2_1$ & $4_1$ & $-0.216339433204765$ &  $0.16860988033085$ &    $0.431009948231322$ &   $-4.62780396148396$ \\
\hline
\end{tabular}
}
\caption{Highest-weight Bethe Ansatz solutions for $N=10$, $M=3$. There are $35$ states with only real roots, $35$ states with one two-string and $5$ three-strings.}
\label{tab:N10M3}
\end{table}

\begin{table*}
{\tiny
\hspace{-12mm}
\begin{tabular}{r|rrrr|rrrr|rrrr|r}
 &\multicolumn{4}{r|}{$2J$}&\multicolumn{4}{r|}{$2I_j$}&\multicolumn{4}{r|}{$\lambda$}&$E_0$\\
 \hline
 \hline
 1. &    $-1$ &  $1$ &   $-3$ &  $3$ &   $-1_1$ &        $1_1$ & $-3_1$ &        $3_1$ & $-0.0982825450251129$ & $0.0982825450251129$ & $-0.338864450851479$ &  $0.338864450851479$ &		    $-6.59220734673867$ \\
 2. &    $-1$ &  $1$ &   $-3$ &  $-5$ &  $-1_1$ &        $1_1$ & $-3_1$ &        $-5_1$ &        $-0.0484346065088534$ & $0.147273674497013$ &  $-0.267004842580402$ &  $-0.708181369437459$ &  $-6.04327937431269$ \\
 3. &    $-1$ &  $1$ &   $3$ &   $-5$ &  $-1_1$ &        $1_1$ & $3_1$ & $-5_1$ &        $-0.0784040591088991$ & $0.114862714354582$ &   $0.358219857377605$ &   $-0.78925147649264$ &   $-5.74616491673258$ \\
 4. &    $-1$ &  $-3$ &  $3$ &   $-5$ &  $-1_1$ &        $-3_1$ &        $3_1$ & $-5_1$ &        $-0.0593071101454481$ & $-0.279010526458747$ &  $0.384540608668618$ &   $-0.73122448092544$ &   $-5.39123610035074$ \\
 5. &    $1$ &   $-3$ &  $3$ &   $-5$ &  $1_1$ & $-3_1$ &        $3_1$ & $-5_1$ &        $0.124662932656233$ &   $-0.289919038880709$ &  $0.370868294394876$ &   $-0.754237921414972$ &  $-5.28049010040769$ \\
 6. &    $-1$ &  $1$ &   $-3$ &  $5$ &   $-1_1$ &        $1_1$ & $-3_1$ &        $5_1$ & $-0.114862714354582$ &  $0.0784040591088991$ &  $-0.358219857377605$ &  $0.78925147649264$ &    $-5.74616491673258$ \\
 7. &    $-1$ &  $1$ &   $3$ &   $5$ &   $-1_1$ &        $1_1$ & $3_1$ & $5_1$ & $-0.147273674497013$ &  $0.0484346065088534$ & $0.267004842580402$ &   $0.708181369437459$ &   $-6.04327937431269$ \\
\end{tabular}
%
%
 39. &   $-1$ &  $-5$ &  $5$ &   $7$ &   $-1_1$ &        $-5_1$ &        \multicolumn{2}{c|}{$2_2$} &    $-0.13104806504265$ &   $-0.903369533501862$ & \multicolumn{2}{c|}{$0.806958457457648\pm 0.518952616044813i$} &       $-2.91449112383954$ \\
 40. &   $1$ &   $-5$ &  $5$ &   $7$ &   $1_1$ & $-5_1$ &        \multicolumn{2}{c|}{$2_2$} &    $0.0457679073574631$ &  $-0.926870273170173$ &  \multicolumn{2}{c|}{$0.771539863055997\pm 0.516523606824117i$} &       $-3.03062952071467$ \\
 41. &   $-1$ &  $5$ &   $5$ &   $7$ &   $-1_1$ &        $5_1$ & \multicolumn{2}{c|}{$2_2$} &    $-0.181053532776174$ &  $0.665228176436887$ &  \multicolumn{2}{c|}{$0.380087206678755\pm 0.500181023771475i$} &       $-3.36254697803928$ \\
 42. &   $1$ &   $5$ &   $5$ &   $7$ &   $1_1$ & $5_1$ & \multicolumn{2}{c|}{$2_2$} &    $0.0127089722401683$ &  $0.66879772886112$ &   \multicolumn{2}{c|}{$0.311850280472639\pm 0.500036029910984i$} &      $-3.62674138044743$ \\
 43. &   $-3$ &  $3$ &   $5$ &   $5$ &   $-3_1$ &        $3_1$ & \multicolumn{2}{c|}{$0_2$} &    $-0.355553222914862$ &  $0.355553222914862$ &  \multicolumn{2}{c|}{$\pm 0.5i$} &       $-3.65662043104711$ \\
 44. &   $-3$ &  $-5$ &  $5$ &   $5$ &   $-3_1$ &        $-5_1$ &        \multicolumn{2}{c|}{$0_2$} &    $-0.310549149731366$ &  $-0.79468977085756$ &   \multicolumn{2}{c|}{$0.318574905734768\pm 0.499979423585762i$} &       $-2.91852179883665$ \\
 45. &   $-3$ &  $5$ &   $-5$ &  $-5$ &  $-3_1$ &        $5_1$ & \multicolumn{2}{c|}{$0_2$} &    $-0.360233082912354$ &  $0.848576145478809$ &   \multicolumn{2}{c|}{$-0.100183195816955\pm 0.499999999566872i$} &      $-2.82207745494737$ \\
\end{tabular}
}
\end{table*}

\begin{table}
{\tiny
\vspace{-10mm}
\hspace{-8mm}
\begin{tabular}{r|rrrr|rrrr|rrrr|r} 
 76. &   $-1$ &  $9$ &  $7$ &   $5$ &    $-2_1$ &        \multicolumn{3}{c|}{$2_3$} & $-0.229951158944146$ &  $0.701127655052017$ &  \multicolumn{2}{c|}{$0.686505224914291\pm 0.988444451913007i$} &   $-2.19098300562505$ \\
 77. &   $3$ &  $9$ & $7$ &   $5$ &      $2_1$ & \multicolumn{3}{c|}{$2_3$} &    $0.12670445684459$ &  $0.615682260875607$ &   \multicolumn{2}{c|}{$0.605806378341122\pm 0.984707473508637i$} &    $-2.44631207308419$ \\
 78. &   $-3$ &  $9$ & $7$ &   $5$ &     $-4_1$ &        \multicolumn{3}{c|}{$2_3$} &    $-0.486092018830097$ &  $0.747279186877141$ &  \multicolumn{2}{c|}{$0.729659118346677\pm 0.990097903104581i$} &    $-1.55364195816963$ \\
 79. &   $5$ &  $9$ &  $7$ &   $5$ &     $4_1$ & \multicolumn{3}{c|}{$2_3$} &    $0.352172387161592$ &   $0.542873778627496$ &  \multicolumn{2}{c|}{$0.537259991040618\pm 0.974848447561115i$} &    $-1.9297275916029$ \\
 80. &   $-5$ &  $9$ &$7$ &   $5$ &      $-6_1$ &        \multicolumn{3}{c|}{$2_3$} &    $-1.04495025932346$ &  $0.818684030062761$ &   \multicolumn{2}{c|}{$0.795754579948182\pm 0.993172728900323i$} &    $-0.874817198903749$ \\
 81. &   $5$ &  $11$ & $9$ &   $3$ &     $6_1$ & \multicolumn{3}{c|}{$2_3$} &    $0.778299777225231$ &   $0.387696439691344$ &  \multicolumn{2}{c|}{$0.384618695188188\pm 1.00250524683233i$} &     $-1.20307037470831$ \\
\hline
	82. &   $7$ &   $5$ &   $5$ &   $9$ &   \multicolumn{4}{c|}{$0_4$} &    \multicolumn{2}{c}{$\pm 1.51235768130767i$} & \multicolumn{2}{c|}{$\pm 0.5i$} &       $-0.509136384638968$ \\
	83. &   $-9$ &  $-9$ &  $-7$ &  $-7$ &  \multicolumn{4}{c|}{$-2_4$} &   \multicolumn{2}{c}{$-0.698286762160898\pm 1.45697124101182i$} &  \multicolumn{2}{c|}{$-0.686425305832397\pm 0.500238343641787i$} &     $-0.45054849432044$ \\
\hline
\end{tabular}
}
\caption{Highest-weight Bethe Ansatz solutions for $N=10$, $M=4$. Real: $15$; one two-string: $45$; one three-string: $21$; one four-string: $3$; two two-strings: $6$.  }
\label{tab:N10M4}
\end{table}

\begin{table}
{\tiny
\begin{tabular}{r|rrrrr|rrrrr|r}
 &\multicolumn{5}{r|}{$2J$}&\multicolumn{5}{r|}{$2I_j$}& $E_0$\\
 \hline
 \hline
 1. &    $0$ &   $-2$ &  $2$ &   $-4$ &  $4$ &   $0_1$ & $-2_1$ &        $2_1$ & $-4_1$ &        $4_1$ & $-7.01544635449204$ \\
 \hline
 2. &    $0$ &   $-2$ &  $2$ &   $4$ &   $6$ &   $0_1$ & $-2_1$ &        $2_1$ & \multicolumn{2}{c|}{$0_2$} &     $-6.27059743540844$ \\
 3. &    $0$ &   $-2$ &  $-4$ &  $4$ &   $6$ &   $0_1$ & $-2_1$ &        $-4_1$ & \multicolumn{2}{c|}{$0_2$} & $-5.26763488483557$ \\
 4. &    $0$ &   $2$ &   $-4$ &  $4$ &   $6$ &   $0_1$ & $2_1$ & $-4_1$ &        \multicolumn{2}{c|}{$0_2$} &       $-5.3862869245099$ \\
 5. &    $-2$ &  $2$ &   $-4$ &  $4$ &   $6$ &   $-2_1$ &        $2_1$ & $-4_1$ &        \multicolumn{2}{c|}{$0_2$} &       $-5.01429947921123$ \\

	6. &    $0$ &   $-2$ &  $4$ &   $-6$ &  $-4$ &  $0_1$ & $-2_1$ &        $4_1$ & \multicolumn{2}{c|}{$0_2$} &      $-5.3862869245099$ \\
	7. &    $0$ &   $2$ &   $4$ &   $-6$ &  $-4$ &  $0_1$ & $2_1$ & $4_1$ & \multicolumn{2}{c|}{$0_2$} &      $-5.26763488483557$ \\
\hline
\end{tabular}
}
\caption{Highest-weight Bethe Ansatz solutions for $N=10$, $M=5$. Real: $1$; one two-string: $10$; one three-string: $14$; one four-string: $7$; one five-string: $1$; two two-strings: $5$; one two- and one three-string: $3$.  This table gives the quantum numbers and energy, the following one 
the rapidities.}
\label{tab:N10M5}
\end{table}

\begin{table}
{\tiny
\begin{tabular}{r|rrrrr|}
 &\multicolumn{5}{r|}{$\lambda$}\\
 \hline
 \hline
 1. & $0$ &   $-0.216201306054532$ &  $0.216201306054532$ &  $-0.598086994989324$ &  $0.598086994989324$ \\
 \hline
 2. &    $0$ &   $-0.236123983535462$ & $0.236123983535459$ &  \multicolumn{2}{c|}{$\pm 0.5i$} \\
 3. &    $-0.0127828374344248$ &  $-0.230841400083837$  & $-0.623492239903063$ & \multicolumn{2}{c|}{$0.433558238710663\pm 0.500715214079799i$} \\
 4. &    $-0.0127083812623862$ & $0.214599696433234$ & $-0.648132093473336$ & \multicolumn{2}{c|}{$0.223120389151246\pm 0.500003412060902i$} \\
 5. &    $-0.234423274285998$ &  $0.207588644710284$ & $-0.633319866543208$ &  \multicolumn{2}{c|}{$0.330077248059462\pm 0.500093388279514i$} \\
6. &    $0.0127083812623863$ &  $-0.214599696433234$ &  $0.648132093473336$ & \multicolumn{2}{c|}{$-0.223120389151246\pm 0.500003412060902i$} \\
	7. &    $0.0127828374344248$ &  $0.230841400083837$ &   $0.623492239903063$ & \multicolumn{2}{c|}{$-0.433558238710663\pm 0.500715214079799i$} \\
\hline
\end{tabular}
}
\caption{Same situation as for Table \ref{tab:N10M5}.  Here, only the rapidities are given, using the same ordering of states.}
\label{tab:N10M5b}
\end{table}

\bibliography{XXX_DEV.bib}

\end{document}